\documentstyle[preprint,prd,aps]{revtex}
 
\begin{document}
\mbox{\hspace*{58ex}} hep-th/9607150 

\mbox{\hspace*{58ex}} OCU-PHYS-160 

\mbox{\hspace*{58ex}} February 1997 

\mbox{\hspace*{58ex}} (Revised version) 

\vspace*{10mm}

\begin{center}
{\Large {\bf 
Screening of mass singularities and finite soft-photon production 
rate in hot QCD }} 

\end{center}

\hspace*{3ex}

\hspace*{3ex}

\begin{center} 
{\large {\sc A. Ni\'{e}gawa}\footnote{ 
E-mail: niegawa@sci.osaka-cu.ac.jp}

{\normalsize\em Department of Physics, Osaka City University } \\ 
{\normalsize\em Sumiyoshi-ku, Osaka 558, Japan} } \\
\end{center} 

\hspace*{2ex}

\hspace*{2ex}
\begin{center} 
{\large {\bf Abstract}} \\ 
\end{center} 
\begin{quotation}
The production rate of a soft photon from a hot quark-gluon plasma 
is computed to leading order at logarithmic accuracy. The canonical 
hard-thermal-loop resummation scheme leads to logarithmically 
divergent production rate due to mass singularities. We show that 
these mass singularities are screened by employing the effective 
hard-quark propagator, which is obtained through resummation of 
one-loop self-energy part in a self-consistent manner. The 
damping-rate part of the effective hard-quark propagator, rather 
than the thermal-mass part, plays the dominant role of screening 
mass singularities. Diagrams including photon--(hard-)quark vertex 
corrections also yield leading contribution to the production rate. 
\end{quotation}
\newpage
\setcounter{equation}{0}
\setcounter{section}{0}
\section{Introduction} 
\def\theequation{\mbox{\arabic{section}.\arabic{equation}}}
It has been established by Pisarski and Braaten and by Frenkel and 
Taylor \cite{pis1,bra1} that, in perturbative thermal QCD, the 
resummations of the leading-order terms, called hard thermal loops, 
are necessary. In thermal massless QCD, we encounter the infrared 
and mass or collinear singularities. The hard-thermal-loop (HTL) 
resummed propagators soften or screen the infrared singularities, 
and render otherwise divergent physical quantities finite 
\cite{tho,bai1}, if they are not sensitive to a further resummation 
of the corrections of $O (g^2 T)$. There are some physical 
quantities which are sensitive to $O (g^2 T)$ corrections, among 
those is the damping rate of a moving particle in a hot quark-gluon 
plasma. Much work has been devoted to this issue 
\cite{pis1,damp,damp1,l-s}. (For reviews of infrared and mass 
singularities in thermal field theory, we refer to \cite{leb1,leb}.) 

Among the thermal reactions, which are expected to serve as 
identifying the hot quark-gluon plasma is the soft-photon 
$(E = O (gT))$ production. This process is analyzed in 
\cite{bai,aur} to leading order within the HTL resummation scheme. 
The conclusion is that the production rate is logarithmically 
divergent, owing to mass singularities. 

The mass singularities found in \cite{bai,aur} arise from bare 
(massless) hard-quark propagators that are on the mass-shell. This 
is a signal \cite{reb,k-r-s,reb1} of necessity of resummation for 
such propagators. Substituting the effective hard-quark propagators, 
$\displaystyle{ \raisebox{1.1ex}{\scriptsize{$\diamond$}}} 
\mbox{\hspace{-0.33ex}} S$s, which is obtained by resumming the 
one-loop self-energy part in a self-consistent manner (cf. e.g. 
\cite{sin-nie}), for the bare propagators $S$s, we show that the 
mass singularities are screened and the diverging factor in the 
production rate turns out to $\ln (g^{- 1})$. This substitution 
violates the current-conservation condition. For recovering it, 
photon-quark vertex corrections should be taken into account. Among 
those is a set of diagrams that yields leading contribution to the 
production rate. 

Here it is worth recording the relations between the differential 
rate $E \, d W / d^{\, 3} p$ of a soft-photon [$P^\mu = (E, 
{\bf p}), \, E = O (g T)$] production, to be analyzed in this 
paper, and other quantities which are of interest in the literature. 
The traditionally defined production rate $\Gamma_p$ is related to 
$E \, d W / d^{\, 3} p$ as 
\begin{equation} 
\Gamma_p = (2 \pi)^3 \, \frac{1}{E} \, \left( E \, 
\frac{d W}{d^{\, 3} p} \right) \, . 
\label{hajimari} 
\end{equation} 
The decay rate, $\Gamma_d$, of a soft photon in a hot quark-gluon 
plasma is related to $\Gamma_p$ as 
\[ 
\Gamma_d = \frac{1}{2} \, \frac{1 + n_B (E)}{n_B (E)} \, 
\Gamma_p \simeq \frac{1}{2} \, \Gamma_p \, , 
\] 
where $n_B (E)$ $=$ $1 / (e^{E / T} - 1)$ $\simeq$ $T / E$ is the 
Bose-distribution function. The damping rate, $\gamma$, of a 
transverse soft photon is related to $\Gamma_p$ as 
\begin{equation} 
\gamma = \frac{1}{4 \, n_B (E)} \, \Gamma_p \simeq \frac{E}{4 \, T} 
\, \Gamma_p \, . 
\label{tsugi} 
\end{equation} 

In Sec. II, we compute the singular contribution to the production 
rate of a soft photon to leading order in real-time thermal field 
theory and reproduce the result in \cite{bai,aur}. In Sec. III, we 
modify the analysis in Sec. II by substituting 
$\displaystyle{ \raisebox{1.1ex}{\scriptsize{$\diamond$}}} 
\mbox{\hspace{-0.33ex}} S$s for $S$s, the $S$s which are responsible 
for mass singularity, and show that mass singularity is screened. 
Then, the contribution to the production rate is evaluated to 
leading order at logarithmic accuracy, by which we mean that the 
factor of $O ( 1 / \ln (g^{- 1}) )$ is ignored when compared to 
the factor of $O (1)$. The contribution thus obtained is gauge 
independent. In Sec. IV, we analyze corrections to the photon-quark 
vertex and then compute the contribution of them to the production 
rate. The resultant contribution coincides with the contribution 
obtained in Sec. III. Sec. V is devoted to discussions and 
conclusions. Appendix A collects formulas used in this paper. 
Appendix B contains calculation of ladder diagrams (cf. Fig. 6 
below) for the photon--hard-quark vertex, in which some of the gluon 
rungs carry hard momenta. In Appendix C, we briefly analyze a class 
of corrections to the photon-quark vertex, which seemingly is of the 
same order of magnitude as the bare photon-quark vertex, and show 
that, eventually, it is nonleading. 

We here introduce notations, $O [g^n T^\ell]$ and $O \{ g^n \}$, 
which we use throughout this paper. 
\begin{description} 
\item $A$ is of $O [g^n T^\ell]$: $A$ is of $O (g^n T^\ell)$, up to 
a possible factor of $\ln (g^{- 1})$. 
\item The contribution $A$ is of $O \{ g^n \}$: The contribution 
$A$ is $O [g^n]$ smaller than the corresponding leading 
contribution. 
\end{description}
\setcounter{equation}{0}
\setcounter{section}{1}
\section{Leading-order calculation in HTL-resummation scheme} 
\def\theequation{\mbox{\arabic{section}.\arabic{equation}}}
The purpose of this section is to compute the differential rate of a 
soft-photon production to lowest nontrivial order within 
HTL-resummation scheme. We work in massless \lq\lq QCD'' with the 
color group $S U (N_c)$ and $N_f$ quarks. 
\subsection{Preliminary} 
After summing over the polarizations of the photon, the differential 
rate of a soft-photon production is given by 
\begin{equation} 
E \, \frac{dW}{d^{\, 3} p} = \frac{i}{2 \, (2 \pi)^3} \, g_{\mu \nu} 
\, \Pi^{\mu \nu}_{12} (P) \, , 
\label{def} 
\end{equation} 
where $P^\mu$ $=$ $(E, {\bf p})$. In (\ref{def}), 
$\Pi^{\mu \nu}_{12}$ is the $(1,2)$ component of the photon 
polarization tensor in the real-time formalism based on the time 
path $C_1 \oplus C_2 \oplus C_3$ in the complex time plane; $C_1 = 
- \infty \to + \infty, \, C_2 = + \infty \to - \infty, \, C_3 = - 
\infty \to - \infty - i / T$. [The time-path segment $C_3$ does not 
play \cite{leb1,lan,nie3} any explicit role in the present context.] 
The fields whose time arguments are lying on $C_1$ and on $C_2$ are 
referred, respectively, to as the type-1 and type-2 fields. A vertex 
of type-1 (type-2) fields is called a type-1 (type-2) vertex. Then 
$\Pi^{\mu \nu}_{12}$ in (\ref{def}) is the \lq\lq thermal vacuum 
polarization between the type-2 photon and the type-1 photon''. 

To leading order, Fig. 1 is the diagram \cite{bai,aur} that 
contributes to $E \, dW /d^{\, 3} p$. In Fig. 1, $p_0 = p = E$ and 
\lq\lq 1'' and \lq\lq 2'' stand for the thermal indices, which 
specify the type of vertex. Fig. 1 leads to 
\begin{eqnarray} 
\Pi^{\mu \nu}_{1 2} (P) & = & - \, i \, e_q^2 e^2 N_c \int 
\frac{d^{\, 4} K}{(2 \pi)^4} \, tr \Big[ \displaystyle{ 
\raisebox{0.6ex}{\scriptsize{*}}} \! S_{i_1 i_4} (K) \nonumber \\ 
& & \times \left( \displaystyle{\raisebox{0.6ex}{\scriptsize{*}}} 
\Gamma^\nu (K, K') \right)^{2}_{i_4 i_3} 
\displaystyle{\raisebox{0.6ex}{\scriptsize{*}}} \! S_{i_3 i_2} (K') 
\nonumber \\ 
& & \times \left( \displaystyle{\raisebox{0.6ex}{\scriptsize{*}}} 
\Gamma^\mu (K', K) \right)^{1}_{i_2 i_1} \Big] \, , 
\label{pi} 
\end{eqnarray}
where $i_1, ... , i_4$ are the thermal indices that specify the 
field type. In (\ref{pi}), all the momenta $P, K$ and $K'$ are soft 
$(\sim gT)$, so that both photon-quark vertices, 
$\displaystyle{\raisebox{0.6ex}{\scriptsize{*}}} 
\Gamma^\nu$ and $\displaystyle{\raisebox{0.6ex}{\scriptsize{*}}} 
\Gamma^\mu$, as well as both quark propagators, $
\displaystyle{\raisebox{0.6ex}{\scriptsize{*}}} \! S_{i_1 i_4}$ and 
$\displaystyle{\raisebox{0.6ex}{\scriptsize{*}}} \! S_{i_3 i_2}$, 
are HTL-resummed effective ones (cf. (\ref{gamma}) - 
(\ref{hito-matome}) below and (\ref{eff1}) - (\ref{eff10}) in 
Appendix A). [Throughout this paper, a capital letter like $P$ 
denotes the four momentum, $P = (p_0, {\bf p})$, and a lower-case 
letter like $p$ denotes the length of the three vector, $p = 
|{\bf p}|$. The unit three vector along the direction of, say, 
${\bf p}$ is denoted as $\hat{{\bf p}} \equiv {\bf p}/p$. The null 
four vector like $\hat{P}_\tau$ $(\tau = \pm)$ is defined as 
$\hat{P}_\tau = (1, \tau \, \hat{{\bf p}})$ and $\hat{P}$ $\equiv$ 
$\hat{P}_{\tau = +}$.] 

As a technical device, we decompose $g_{\mu \nu}$ in (\ref{def}) 
into two parts as 
\begin{eqnarray} 
g_{\mu \nu} & = & g^{(t)}_{\mu \nu} (\hat{{\bf p}}) + 
g^{(\ell)}_{\mu \nu} (\hat{P}) \, , 
\label{g0} \\ 
g^{(t)}_{\mu \nu} (\hat{{\bf p}}) & \equiv & - \sum_{i, \, j = 1}^3 
g_{\mu i} \, g_{\nu j} \, \left( \delta_{i j} - \hat{p}_i \hat{p}_j 
\right) \, , 
\label{gt} \\ 
g_{\mu \nu}^{(\ell)} (\hat{P}) & \equiv & g_{\mu 0} \hat{P}_\nu + 
g_{\nu 0} \hat{P}_\mu - \hat{P}_\mu \hat{P}_\nu \, . 
\label{g} 
\end{eqnarray} 
Substituting (\ref{g0}) for $g_{\mu \nu}$ in (\ref{def}), we have, 
with an obvious notation, 
\begin{equation} 
E \, \frac{dW}{d^{\, 3} p} = E \, \frac{dW^{(t)}}{d^{\, 3} p} + 
E \, \frac{dW^{(\ell)}}{d^{\, 3} p} \, . 
\label{bunnkatsu} 
\end{equation} 
Now we observe that $\left( 
\displaystyle{\raisebox{0.6ex}{\scriptsize{*}}} 
\Gamma^\mu \right)^1_{j i}$ in (\ref{pi}) is written as 
\begin{eqnarray}
\left( \displaystyle{\raisebox{0.6ex}{\scriptsize{*}}} \Gamma^\mu 
(K', K) \right)^1_{j i} & = & \gamma^\mu \, \delta_{1 j} \, 
\delta_{1 i} + \left( 
\displaystyle{\raisebox{0.6ex}{\scriptsize{*}}} \tilde{\Gamma}^\mu 
(K', K) \right)^1_{j i} \, , 
\label{gamma} \\ 
\left( \displaystyle{\raisebox{0.6ex}{\scriptsize{*}}} 
\tilde{\Gamma}^\mu (K', K) \right)^1_{j i} & = & \frac{m_f^2}{4 \pi} 
\int d \Omega \, \hat{Q}^\mu 
\hat{{Q\kern-0.1em\raise0.3ex\llap{/}\kern0.15em\relax}} \, 
f_{j i} (\hat{Q}, K', K) \, , 
\label{gamma-star} 
\end{eqnarray}
where $m_f$, thermally induced quark mass, is defined as 
\begin{equation} 
m_f^2 = \frac{\pi \alpha_s}{2} \, C_F \, T^2 \;\;\;\;\;\;\;\;\; 
\left( C_F = \frac{N_c^2 - 1}{2 N_c} \right) 
\label{Mf} 
\end{equation} 
and 
\begin{mathletters} 
\label{hito-matome} 
\begin{eqnarray} 
f_{1 1} & = & \frac{{\cal P}}{K' \cdot \hat{Q}} \, \frac{{\cal P}}{K 
\cdot \hat{Q}} \, , \eqnum{2.10a} \\ 
f_{2 1} & = & i \pi \frac{{\cal P}}{K \cdot \hat{Q}} \, \delta 
\left( K' \cdot \hat{Q} \right) = i \pi \frac{{\cal P}}{P \cdot 
\hat{Q}} \, \delta \left( K' \cdot \hat{Q} \right) \, , \nonumber \\ 
\eqnum{2.10b} \\ 
f_{1 2} & = & - i \pi \frac{{\cal P}}{K' \cdot \hat{Q}} \, \delta 
\left( K \cdot \hat{Q} \right) = i \pi \frac{{\cal P}}{P \cdot 
\hat{Q}} \, \delta \left( K \cdot \hat{Q} \right) \, , \nonumber \\ 
\eqnum{2.10c} \\ 
f_{2 2} & = & \pi^2 \, \delta \left( K' \cdot \hat{Q} \right) \, 
\delta \left( K \cdot \hat{Q} \right) \, . 
\eqnum{2.10d} 
\end{eqnarray} 
\end{mathletters} 

\noindent In (\ref{hito-matome}), ${\cal P}$ indicates the 
principal-value prescription. $\left( 
\displaystyle{\raisebox{0.6ex}{\scriptsize{*}}} \tilde{\Gamma}^\mu 
\right)^1_{j i}$, Eq. (\ref{gamma-star}), is the HTL contribution, 
in terms of an angular integral, and $Q^\mu = q \hat{Q}^\mu$ is the 
hard momentum circulating along the HTL. $\left( 
\displaystyle{\raisebox{0.6ex}{\scriptsize{*}}} \Gamma^\nu (K, K') 
\right)^2_{j i}$ in (\ref{pi}) is obtained from 
$\left( 
\displaystyle{\raisebox{0.6ex}{\scriptsize{*}}} \Gamma^\mu (K', K) 
\right)^1_{j i}$ through the relation 
\begin{equation} 
\left( \displaystyle{\raisebox{0.6ex}{\scriptsize{*}}} \Gamma^\nu 
(K, K') \right)^2_{j i} = - \gamma_0 \left[ 
\left( \displaystyle{\raisebox{0.6ex}{\scriptsize{*}}} \Gamma^\nu 
(K', K) \right)^1_{\underline{i} \underline{j}} \right]^\dagger 
\gamma_0 \, , 
\label{recip} 
\end{equation} 
where $\underline{i}$ $=$ $2$ for $i = 1$ and $\underline{i}$ $=$ 
$1$ for $i = 2$. Note that (\ref{recip}) is the general relation, to 
which the photon-quark vertex function is subjected. It turns out 
that the production rate $E \, dW/d^{\, 3} p$ diverges due to mass 
singularities. As in \cite{bai,aur}, we are only interested in the 
divergent parts neglecting all finite contributions. 
\subsection{Computation of $E \, dW^{(t)}/d^{\, 3} p$ in 
(\ref{bunnkatsu})} 
Mass singularities arise from the factor $1 / P \cdot \hat{Q}$ $=$ 
$\{ E (1 - \hat{{\bf p}} \cdot \hat{{\bf q}}) \}^{- 1}$ in 
(\ref{hito-matome}), which diverges at $\hat{{\bf p}} \parallel 
\hat{{\bf q}}$. Let us see the numerator factors in the integrand of 
$E \, dW^{(t)}/d^{\, 3} p$, which are obtained after taking the 
trace of Dirac matrices under the HTL approximation (cf. 
(\ref{pi})). 

\begin{itemize}
\item 
One of the photon-quark vertices in Fig. 1 is the HTL contribution 
and the other is the bare vertex. 

In $E \, d W^{(t)} / d^{\, 3} p$, $\hat{Q}^\mu$ in 
(\ref{gamma-star}) is to be multiplied by $g^{(t)}_{\mu \nu} 
(\hat{{\bf p}})$: $\hat{Q}^\mu g^{(t)}_{\mu \nu} (\hat{{\bf p}})$ 
$=$ $g_{\nu i} [\hat{q}^i - (\hat{{\bf q}} \cdot \hat{{\bf p}}) \, 
\hat{p}^i]$, which vanishes at ${\bf q} \parallel {\bf p}$. Then, 
there is no singularity in the integrand. 
\item 
Both photon-quark ver\-tices in Fig. 1 are the HTL contributions. 

Using (\ref{pi}) and (\ref{gamma-star}), we see that $E \, d W^{(t)} 
/ d^{\, 3} p$ includes $g^{(t)}_{\mu \nu} (\hat{{\bf p}}) 
\hat{Q}^\mu \hat{Q}'^\nu$ $=$ $- \hat{{\bf q}} \cdot 
\hat{{\bf q}}'$ $+$ $(\hat{{\bf q}} \cdot \hat{{\bf p}}) 
(\hat{{\bf p}} \cdot \hat{{\bf q}}')$, where $Q'^\mu$ $=$ $q' 
\hat{Q}'^\mu$ is the hard momentum in $\left( 
\displaystyle{\raisebox{0.6ex}{\scriptsize{*}}} \Gamma^\nu 
\right)^2_{j i}$ in (\ref{pi}).  For $\hat{{\bf q}}$ $=$ 
$\hat{{\bf p}}$ {\em and} $\hat{{\bf q}}'$ $\neq$ $\hat{{\bf p}}$, 
or for $\hat{{\bf q}}'$ $=$ $\hat{{\bf p}}$ {\em and} $\hat{{\bf q}} 
\neq \hat{{\bf p}}$, $g^{(t)}_{\mu \nu} (\hat{{\bf p}}) 
\hat{Q}^\mu \hat{Q}'^\nu$ vanishes. For $\hat{{\bf q}}$ $\simeq$ 
$\hat{{\bf p}}$ {\em and} $\hat{{\bf q}}'$ $\simeq$ $\hat{{\bf p}}$, 
$g^{(t)}_{\mu \nu} (\hat{{\bf p}}) \hat{Q}^\mu \hat{Q}'^\nu$ 
$\propto$ $(1 - \hat{{\bf p}} \cdot \hat{{\bf q}})^{1 / 2}$ $(1 - 
\hat{{\bf p}} \cdot \hat{{\bf q}}')^{1 / 2}$, and the integrations 
over the directions of $\hat{{\bf q}}$ and $\hat{{\bf q}}'$ 
converge. 
\end{itemize}
Thus, $E \, d W^{(t)} / d^{\, 3} p$ is free from singularity. 
\subsection{Computation of $E \, dW^{(\ell)} / d^{\, 3} p$ in 
(\ref{bunnkatsu})} 
Substituting $g^{(\ell)}_{\mu \nu} (\hat{P})$, Eq. (\ref{g}), for 
$g_{\mu \nu}$ in (\ref{def}), we obtain 
\begin{eqnarray} 
E \, \frac{d W^{(\ell)}}{d^{\, 3} p} & = & \frac{e_q^2 e^2 N_c}{2 
(2 \pi)^3} \left[ g_{\mu 0} \hat{P}_\nu + g_{\nu 0} \hat{P}_\mu 
- \hat{P}_\mu \hat{P}_\nu \right] \int \frac{d^{\, 4} K}{(2 \pi)^4} 
\, tr \Big[ \displaystyle{ \raisebox{0.6ex}{\scriptsize{*}}} \! 
S_{i_1 i_4} (K) \nonumber \\ 
& & \times \left( \displaystyle{\raisebox{0.6ex}{\scriptsize{*}}} 
\Gamma^\nu (K, K') \right)^{2}_{i_4 i_3} 
\displaystyle{\raisebox{0.6ex}{\scriptsize{*}}} \! S_{i_3 i_2} (K') 
\left( \displaystyle{\raisebox{0.6ex}{\scriptsize{*}}} \Gamma^\mu 
(K', K) \right)^{1}_{i_2 i_1} \Big] \, . 
\label{pil1} 
\end{eqnarray} 

\noindent It is convenient to employ here the Ward-Takahashi 
relations, 
\begin{equation} 
(K - K')_\mu \displaystyle{ \raisebox{0.6ex}{\scriptsize{*}}} \! 
S_{j i_2} (K') \left( \displaystyle{ 
\raisebox{0.6ex}{\scriptsize{*}}} \Gamma^\mu (K', K) 
\right)^\ell_{i_2 i_1} \displaystyle{ 
\raisebox{0.6ex}{\scriptsize{*}}} \! S_{i_1 i} (K) = \delta_{\ell i} 
\displaystyle{\raisebox{0.6ex}{\scriptsize{*}}} \! S_{j i} (K') - 
\delta_{\ell j} \displaystyle{\raisebox{0.6ex}{\scriptsize{*}}} \! 
S_{j i} (K) \, . 
\label{ward} 
\end{equation} 

\noindent On the R.H.S., no summations are taken over $i$ and $j$. 
Using (\ref{ward}), we can easily see that the term with 
$\hat{P}_\mu \hat{P}_\nu$ in (\ref{pil1}) does not yield 
mass-singular contribution. We then obtain, for the singular 
contributions, 
\begin{eqnarray} 
E \, \frac{d W^{(\ell)}}{d^{\, 3} p} & \simeq & 
\frac{e_q^2 e^2 N_c}{2 (2 \pi)^3} \, \frac{1}{E} \int \frac{d^{\, 4} 
K}{(2 \pi)^4} \left\{ tr \left[ \displaystyle{ 
\raisebox{0.6ex}{\scriptsize{*}}} \! S_{2 i} (K') \left( 
\displaystyle{\raisebox{0.6ex}{\scriptsize{*}}} \Gamma^0 (K', K) 
\right)^{1}_{i 2} - \left( 
\displaystyle{\raisebox{0.6ex}{\scriptsize{*}}} \Gamma^0 (K', K) 
\right)^{1}_{2 i} \displaystyle{ \raisebox{0.6ex}{\scriptsize{*}}} 
\! S_{i 2} (K) \right] \right. \nonumber \\ 
& & \left. + tr \left[ \left( 
\displaystyle{\raisebox{0.6ex}{\scriptsize{*}}} \Gamma^0 (K, K') 
\right)^{2}_{1 i} \displaystyle{ \raisebox{0.6ex}{\scriptsize{*}}} 
\! S_{i 1} (K') - \displaystyle{ \raisebox{0.6ex}{\scriptsize{*}}} 
\! S_{1 i} (K) \left( 
\displaystyle{\raisebox{0.6ex}{\scriptsize{*}}} \Gamma^0 (K, K') 
\right)^{2}_{i 1} \right] \right\} \nonumber \\ 
& = & \frac{e_q^2 e^2 N_c}{(2 \pi)^3} \, \frac{1}{E} \, Re \int 
\frac{d^{\, 4} K}{(2 \pi)^4} \, tr \left[ \displaystyle{ 
\raisebox{0.6ex}{\scriptsize{*}}} \! S_{2 i} (K') \left( 
\displaystyle{\raisebox{0.6ex}{\scriptsize{*}}} \Gamma^0 (K', K) 
\right)^{1}_{i 2} \right. \nonumber \\ 
& & \left. - \left( \displaystyle{ 
\raisebox{0.6ex}{\scriptsize{*}}} \Gamma^0 (K', K) 
\right)^{1}_{2 i} \displaystyle{ \raisebox{0.6ex}{\scriptsize{*}}} 
\! S_{i 2} (K) \right] \, , 
\label{pil22} 
\end{eqnarray} 

\noindent where and in the following in this section the symbol 
\lq\lq $\simeq$'' is used to denote an approximation that is valid 
for keeping the singular contributions. The symbol \lq\lq $Re$'' in 
(\ref{pil22}) means to take the real part of the quantity placed on 
the right of \lq\lq $Re$''. In obtaining (\ref{pil22}), use has been 
made of the relation (\ref{recip}) and 
\[ 
\displaystyle{ \raisebox{0.6ex}{\scriptsize{*}}} \! S_{j i} (K) = 
- \gamma_0 \left[ \displaystyle{ \raisebox{0.6ex}{\scriptsize{*}}} 
\! S_{\underline{i} \underline{j}} (K) \right]^\dagger \gamma_0 
\, . 
\] 
The mass-singular contributions arise from the HTL parts $\left( 
\displaystyle{\raisebox{0.6ex}{\scriptsize{*}}} \tilde{\Gamma}^0 
\right)_{i j}^1$ with $i \neq j$ of 
$\displaystyle{\raisebox{0.6ex}{\scriptsize{*}}} \Gamma^0$s 
in (\ref{pil22}). We use dimensional regularization as defined in 
\cite{bai}, which gives 
\[ 
\int d \Omega \frac{{\cal P}}{P \cdot \hat{Q}} = \frac{2\pi}{E 
\hat{\epsilon}} \, ,  
\] 
where $\hat{\epsilon}$ $=$ $(D - 4) / 2$ with $D$ the space-time 
dimension. Then, from (\ref{gamma-star}) and (\ref{hito-matome}), we 
see that the singular contributions come from the point ${\bf p} 
\parallel {\bf q}$: 
\begin{mathletters} 
\label{HTL-vertex} 
\begin{eqnarray} 
\left( \displaystyle{\raisebox{0.6ex}{\scriptsize{*}}} \Gamma^\mu 
(K', K) \right)^1_{2 1} & = & i \frac{\pi}{2} \frac{m_f^2}{E} 
\frac{1}{\hat{\epsilon}} \hat{P}^\mu 
\hat{{P\kern-0.1em\raise0.3ex\llap{/}\kern0.15em\relax}} \, \delta 
(K' \cdot \hat{P}) \, , 
\eqnum{2.15a} \\ 
\left( \displaystyle{\raisebox{0.6ex}{\scriptsize{*}}} \Gamma^\mu 
(K', K) \right)^1_{1 2} & = & i \frac{\pi}{2} \frac{m_f^2}{E} 
\frac{1}{\hat{\epsilon}} \hat{P}^\mu 
\hat{{P\kern-0.1em\raise0.3ex\llap{/}\kern0.15em\relax}} \, \delta 
(K \cdot \hat{P}) \, . 
\eqnum{2.15b} 
\end{eqnarray} 
\end{mathletters} 
Substituting (\ref{HTL-vertex}), (\ref{eff1}), and (\ref{eff4}) in 
Appendix A into (\ref{pil22}), we obtain for the singular 
contribution, 
\begin{eqnarray} 
E \, \frac{d W^{(\ell)}}{d^{\, 3} p} & \simeq & \frac{e_q^2 \, e^2 
\, N_c}{8 \pi^2} \, \frac{m_f^2}{E} \, \frac{1}{\hat{\epsilon}} \int 
\frac{d^{\, 4} K}{(2 \pi)^3} \, \delta (P \cdot K) \nonumber \\ 
& & \times \sum_{\sigma = \pm} \left( \hat{K}_\sigma \cdot \hat{P} 
\right) \rho_\sigma (K) \, , 
\label{bare-final} 
\end{eqnarray} 
where $\hat{K}_\sigma$ $=$ $(1, \sigma {\bf k})$ and the spectral 
function $\rho_\sigma (K)$ is defined in (\ref{eff10}) in Appendix 
A. In deriving (\ref{bare-final}), use has been made of $n_F (- 
k_0')$ $\simeq$ $n_F (k_0)$ $\simeq$ $1 / 2$, where $n_F (x)$ 
$\equiv$ $1 / (e^{x / T} + 1)$ is the Fermi-distribution function. 
Thus, we have reproduced the result reported in \cite{bai,aur}. 

We encounter the same integral as in (\ref{bare-final}) in the 
hard-photon production case \cite{bai1}. Here we recall that $K$ is 
soft $\sim O (g T)$. Then, the upper limit $k^*$ of the integration 
over $k$ is in the range, $g T << k^* << T$. Referring to 
\cite{bai1}, we have 
\begin{equation} 
E \, \frac{dW}{d^{\, 3} p} \simeq \frac{e_q^2 \, \alpha \, 
\alpha_s}{2 \pi^2} \, T^2 \left( \frac{m_f}{E} \right)^2 
\frac{1}{\hat{\epsilon}} \, \ln \left( \frac{k^*}{m_f} \right) \, . 
\label{w2} 
\end{equation} 

The hard contribution should be added to the soft contribution 
(\ref{w2}). Besides a factor of $\{ \ln (T / k^*) + O (1) \}$, the 
mechanism of arising mass singularity in the former is the same as 
in the latter \cite{reb1}. Thus we finally obtain 
\[ 
E \, \frac{dW}{d^{\, 3} p} \simeq \frac{e_q^2 \, \alpha \, 
\alpha_s}{2 \pi^2} \, T^2 \left( \frac{m_f}{E} \right)^2 
\frac{1}{\hat{\epsilon}} \, \ln \left( \frac{T}{m_f} \right) \, . 
\] 
\setcounter{equation}{0}
\setcounter{section}{2}
\section{Modified hard-quark propagators and screening of 
mass-singularity} 
\def\theequation{\mbox{\arabic{section}.\arabic{equation}}}
\subsection{Preliminary} 
In Sec. II, we have seen that the singular contribution comes from 
the region $\hat{P} \cdot \hat{Q}$ $=$ $1 - \hat{{\bf p}} \cdot 
\hat{{\bf q}}$ $\simeq$ $0$ in $\left( 
\displaystyle{\raisebox{0.6ex}{\scriptsize{*}}} \tilde{\Gamma}^\mu 
\right)^1_{j i}$ with $j \neq i$ and $\hat{P} \cdot \hat{Q}'$ $=$ 
$1 - \hat{{\bf p}} \cdot \hat{{\bf q}}'$ $\simeq$ $0$ in $\left( 
\displaystyle{\raisebox{0.6ex}{\scriptsize{*}}} \tilde{\Gamma}^\nu 
\right)^2_{j i}$ with $j \neq i$. Let us first see how does the 
factor ${\cal P} / \hat{P} \cdot \hat{Q}$, which develops 
singularity, come about in these 
$\displaystyle{\raisebox{0.6ex}{\scriptsize{*}}} \tilde{\Gamma}$s 
(cf. (\ref{gamma-star}) with (\ref{hito-matome})). 

The diagram to be analyzed is depicted in Fig. 2, where $Q^\mu$ is 
hard while $P$, $K$, and $K'$ are soft. $\displaystyle{ 
\raisebox{0.6ex}{\scriptsize{*}}} \tilde{\Gamma}^\mu$ computed 
within the canonical HTL-resummation scheme is gauge independent, 
which diverges (cf. (\ref{HTL-vertex})). The gauge-parameter 
dependent part of the hard-gluon propagator (cf. (\ref{bosb1}) in 
Appendix A) leads to nonleading contribution. As mentioned in Sec. 
II, throughout this paper, we pursue the leading contribution that 
diverges and ignore finite as well as nonleading contributions. 
Then, we can use Feynman gauge for the gluon propagator in Fig. 2: 
\begin{eqnarray} 
& & \left( \displaystyle{\raisebox{0.6ex}{\scriptsize{*}}} 
\tilde{\Gamma}^\mu (K', K) \right)^\ell_{i j} \nonumber \\ 
& & \mbox{\hspace*{5ex}} = i \, (-)^{i + j + \ell} g^2 \, C_F \, 
g^{\rho \sigma} \int \frac{d^{\, 4} Q}{(2 \pi)^4} \, \gamma_\rho \, 
S_{i \ell} (Q + K') \nonumber \\ 
& & \mbox{\hspace*{7.5ex}} \times \gamma^\mu \, S_{\ell j} (Q + K) 
\, \gamma_\sigma \, \Delta_{j i} (Q) \, , 
\label{shuppatsu} 
\end{eqnarray} 
with no summation over $\ell, i$, and $j$. In (\ref{shuppatsu}), 
$S_{\ell j}$ is the bare thermal quark propagator and $- g^{\rho 
\sigma} \Delta_{j i}$ is the bare thermal gluon propagator (cf. 
Appendix A). It is worth remarking that $\displaystyle{ 
\raisebox{0.6ex}{\scriptsize{*}}} \tilde{\Gamma}^\mu$s in 
(\ref{shuppatsu}) satisfy the relation (\ref{recip}). 

Substituting (\ref{a-1}) in Appendix A, we obtain for the leading 
contribution, 
\begin{eqnarray} 
& & \left( \displaystyle{\raisebox{0.6ex}{\scriptsize{*}}} 
\tilde{\Gamma}^\mu (K', K) \right)^\ell_{i j} \nonumber \\ 
& & \mbox{\hspace*{5ex}} \simeq - 4 i (-)^{i + j + \ell} g^2 \, C_F 
\int \frac{d^{\, 4} Q}{(2 \pi)^4} \sum_{\tau = \pm} \hat{Q}^\mu_\tau 
\nonumber \\ 
& & \mbox{\hspace*{8ex}} \times 
\hat{{Q\kern-0.1em\raise0.3ex\llap{/}\kern0.15em\relax}}_\tau \, 
\tilde{S}^{(\tau)}_{i \ell} (Q + K') \, \tilde{S}^{(\tau)}_{\ell j} 
(Q + K) \, \Delta_{j i} (Q) \, . 
\label{ver-1} 
\end{eqnarray} 
Using (\ref{a-2}) and (\ref{a-3}) (in Appendix A) for 
$\tilde{S}^{(\tau)}$s, we see that in $\left( 
\displaystyle{\raisebox{0.6ex}{\scriptsize{*}}} \tilde{\Gamma}^\mu 
\right)^1_{2 1}$, for example, the singular contribution comes from 
\begin{mathletters} 
\label{ver-2} 
\begin{eqnarray} 
& & \tilde{S}^{(\tau)}_{2 1} (Q + K') \, Re \tilde{S}^{(\tau)}_{1 1} 
(Q + K) \nonumber \\ 
& & \mbox{\hspace*{5ex}} = - i \frac{\pi}{2} \, n_F (- q_0) \, 
\delta (q_0 + k_0' - \tau |{\bf q} + {\bf k}'|) \nonumber \\ 
& & \mbox{\hspace*{8ex}} \times \frac{{\cal P}}{q_0 + k_0 - \tau 
|{\bf q} + {\bf k}|} 
\eqnum{3.3a} \\ 
& & \mbox{\hspace*{5ex}} \simeq - i \frac{\pi}{2} \, n_F (- \tau q) 
\, \delta (q_0 - \tau q + K' \cdot \hat{Q}_\tau ) \nonumber \\ 
& & \mbox{\hspace*{8ex}} \times \frac{{\cal P}}{q_0 - \tau q + K' 
\cdot \hat{Q}_\tau} 
\eqnum{3.3b} \\ 
& & \mbox{\hspace*{5ex}} = - i \frac{\pi}{2} [ \theta (\tau) - \tau 
n_F (q) ] \nonumber \\ 
& & \mbox{\hspace*{8ex}} \times \delta (q_0 - \tau q + K' \cdot 
\hat{Q}_\tau ) \, \frac{1}{P \cdot \hat{Q}_\tau} \, . 
\eqnum{3.3c} 
\end{eqnarray} 
\end{mathletters} 

\noindent Since $P \cdot \hat{Q}_\tau$ $\geq$ $0$, the ${\cal P}$ 
prescription is dropped in the last line. For the $\tau$ $=$ $-$ 
sector, the integration variable ${\bf q}$ in (\ref{ver-1}) is 
changed to $- {\bf q}$, so that $P \cdot \hat{Q}_-$ $\to$ $P \cdot 
\hat{Q}_+$ $=$ $P \cdot \hat{Q}$. Carrying out the integration over 
$q_0$ and then over $q = |{\bf q}|$, we obtain (\ref{gamma-star}) 
with (2.10b). In (3.3c), a singularity appears at $P \cdot 
\hat{Q}_\tau$ $=$ $0$, i.e. $\hat{{\bf q}}$ $=$ $\tau 
\hat{{\bf p}}$. 

It can readily be seen that this singularity is not the artifact of 
the approximation made at (3.3b). In order to see this, consider the 
process \lq\lq $q$'' $(Q + K)$ $\to$ $q$ $(Q + K')$ $+$ $\gamma 
(P)$, where $q (Q + K')$ [$\gamma (P)$] is the on-shell quark 
[photon] and \lq\lq $q$'' is the off-shell quark (cf. Fig. 2). The 
propagator $1 / (Q + K)^2$ has singularity at ${\bf p} \parallel 
({\bf q} + {\bf k}')$. In fact, $1 / (Q + K)^2$ $=$ $1 / (Q + K' + 
P)^2$ $=$ $[2 \{ p |{\bf q} + {\bf k}'|$ $- {\bf p} \cdot ({\bf q} + 
{\bf k}') \} ]^{- 1}$ $=$ $\infty$, the collinear or mass 
singularity. As in the present example, mass singularity may emerge 
\cite{kin} from the small phase-space region where the momenta $R$s 
(being kinematically constrained to $R^2 \geq 0$ or $R^2 \leq 0$) 
carried by bare propagators are close to the mass-shell, $R^2 \simeq 
0$. Other parts of the diagram under consideration do not 
participate directly in the game. As seen above, relevant parts in 
(\ref{shuppatsu}) with $\ell = 1$, $i = 2$, and $j = 1$ are $S_{2 1} 
(Q + K')$, $S_{1 1} (Q + K)$, and the external photon line. We 
assume that the photon is not thermalized, so that no (thermal) 
correction to the on-shell photon should be taken into account. 

Similar observation may be made for $\left( 
\displaystyle{\raisebox{0.6ex}{\scriptsize{*}}} \tilde{\Gamma}^\mu 
\right)^1_{1 2}$. (Note that $\left( 
\displaystyle{\raisebox{0.6ex}{\scriptsize{*}}} \tilde{\Gamma}^\mu 
\right)^1_{1 1}$ and $\left( 
\displaystyle{\raisebox{0.6ex}{\scriptsize{*}}} \tilde{\Gamma}^\mu 
\right)^1_{2 2}$ have not yielded mass-sin\-gu\-lar contribution.) 
\subsection{Modified hard-quark propagator} 
Above observation leads us to look into the hard-quark propagator 
close to the light-cone through the analysis of one-loop thermal 
self-energy part $\tilde{\Sigma}_F (R)$ as depicted in Fig. 3. 
Here $\tilde{\Sigma}_F (R)$ is the quasiparticle or diagonalized 
self-energy part \cite{lan}. When $Q$ ($R - Q$) is soft, the 
effective gluon (quark) propagator, $\displaystyle{ 
\raisebox{0.6ex}{\scriptsize{*}}} \! \Delta^{\rho \sigma} (Q)$ 
$(\displaystyle{ \raisebox{0.6ex}{\scriptsize{*}}} \! S (R - Q))$, 
should be assigned to the gluon (quark) line in Fig. 3. We are 
aiming at constructing the $\tilde{\Sigma}_F (R)$-resummed 
hard-quark propagator $\displaystyle{ 
\raisebox{1.1ex}{\scriptsize{$\diamond$}}} \mbox{\hspace{-0.33ex}} 
S_F (R)$, from which $\displaystyle{ 
\raisebox{1.1ex}{\scriptsize{$\diamond$}}} \mbox{\hspace{-0.33ex}} 
S_{i j} (R)$ $(i, j = 1, 2)$ is obtained through standard manner 
\cite{lan}. When $R^2 \simeq 0$, $Im \, \tilde{\Sigma}_F (R)$ is 
sensitive \cite{damp,damp1,l-s,sin-nie} to the region where $Q^2$ is 
soft, $R$ is hard, and $(R - Q)^2 \simeq 0$. Then, in contrast to 
the case of soft momentum, in determining $\tilde{\Sigma}_F (R)$ 
with hard $R$ with $R^2 \simeq 0$, we need a knowledge of 
$\displaystyle{ \raisebox{1.1ex}{\scriptsize{$\diamond$}}} 
\mbox{\hspace{-0.33ex}} S_F (R - Q)$ with $(R - Q)^2 \simeq 0$. 
Thus, we should determine $\tilde{\Sigma}_F (R)$ in a 
self-consistent manner. (See, e.g., \cite{damp1,sin-nie}.) This is 
also the case for hard-gluon propagator, $\displaystyle{ 
\raisebox{1.1ex}{\scriptsize{$\diamond$}}} \mbox{\hspace{-0.33ex}} 
\Delta_F^{\mu \nu} (Q)$, and hard--FP-ghost propagator, 
$\displaystyle{\raisebox{1.1ex}{\scriptsize{$\diamond$}}} 
\mbox{\hspace{-0.33ex}} \Delta_F^{(F P)} (Q)$. 

A self-consistent determination of $\tilde{\Sigma}_F (R)$ and 
$\displaystyle{ \raisebox{1.1ex}{\scriptsize{$\diamond$}}} 
\mbox{\hspace{-0.33ex}} S (R)$s (as well as of $\displaystyle{ 
\raisebox{1.1ex}{\scriptsize{$\diamond$}}} \mbox{\hspace{-0.33ex}} 
\Delta^{\mu \nu}$s and $\displaystyle{ 
\raisebox{1.1ex}{\scriptsize{$\diamond$}}} \mbox{\hspace{-0.33ex}} 
\Delta^{(F P)}$s) is carried out in \cite{sin-nie}. Here we 
summarize the result. $\tilde{\Sigma}_F (R)$ takes the form 
\begin{equation} 
\tilde{\Sigma}_F (R) \simeq \epsilon (r_0) \left[ \frac{m_f^2}{r} - 
i \gamma_q \right] \, \gamma^0 + O (g^2) \times 
{R\kern-0.1em\raise0.3ex\llap{/}\kern0.15em\relax} \, , 
\label{yama-1} 
\end{equation} 
where $m_f$ is as in (\ref{Mf}) and 
\begin{equation} 
\gamma_q = \frac{g^2}{4 \pi} \, C_F \, T \, \ln (g^{- 1}) \left[ 1 + 
O \left( \frac{\ln \{ \ln (g^{- 1})\}}{\ln (g^{- 1})} \right) 
\right] \, . 
\label{yama-2} 
\end{equation} 
As in \cite{damp,damp1,l-s,sin-nie}, the form (\ref{yama-2}) 
is valid at logarithmic accuracy, i.e., the term of $O (g^2 T)$ is 
ignored when compared to the term of $O (g^2 T \, \ln (g^{- 1}))$. 
It is to be noted that $\gamma_q$ in (\ref{yama-2}) is independent 
of (hard-)$Q$. If necessary, one may explicitly evaluate the term $O 
(\ln ( \ln (g^{- 1}) ) / \ln (g^{-1}))$. 

The (part of the) term $O (g^2) \times 
{R\kern-0.1em\raise0.3ex\llap{/}\kern0.15em\relax}$ in 
(\ref{yama-1}) is absorbed into the wave-function renormalization 
constant. The remainder, which depends on the renormalization 
scheme, gives overall correction of $O (g^2)$ to 
$\displaystyle{ \raisebox{1.1ex}{\scriptsize{$\diamond$}}} 
\mbox{\hspace{-0.33ex}} S_F (R)$ and does not affect the structure 
of $\displaystyle{ \raisebox{1.1ex}{\scriptsize{$\diamond$}}} 
\mbox{\hspace{-0.33ex}} S_F (R)$ at the region of our interest. 
Then, the term $O (g^2) \times 
{R\kern-0.1em\raise0.3ex\llap{/}\kern0.15em\relax}$ in 
(\ref{yama-1}) leads to the contribution of $O \{ g^2 \}$ to the 
soft-photon production rate, and we ignore it in the following. 
(\lq\lq $O \{ g^2 \}$'' is defined at the end of Sec. I.)  The rest 
of the term in $\tilde{\Sigma}_F (R)$ is gauge independent. 

$\tilde{\Sigma}_F (R)$-resummed propagators 
$\displaystyle{ \raisebox{1.1ex}{\scriptsize{$\diamond$}}} 
\mbox{\hspace{-0.33ex}} S_F (R)$s can be written as 
\begin{eqnarray} 
\displaystyle{ \raisebox{1.1ex}{\scriptsize{$\diamond$}}} 
\mbox{\hspace{-0.33ex}} S_{j \ell} (R) & = & \sum_{\tau = \pm} 
\hat{{R\kern-0.1em\raise0.3ex\llap{/}\kern0.15em\relax}}_\tau 
\displaystyle{ \raisebox{1.1ex}{\scriptsize{$\diamond$}}} 
\mbox{\hspace{-0.33ex}} \tilde{S}_{j \ell}^{(\tau)} (R) 
\;\;\;\;\;\;\;\;\; (j, \, \ell = 1, 2) \, , 
\\ 
\displaystyle{ \raisebox{1.1ex}{\scriptsize{$\diamond$}}} 
\mbox{\hspace{-0.33ex}} \tilde{S}_{1 1}^{(\tau)} (R) & \simeq & - 
\left[ \displaystyle{ \raisebox{1.1ex}{\scriptsize{$\diamond$}}} 
\mbox{\hspace{-0.33ex}} \tilde{S}_{2 2}^{(\tau)} (R) \right]^* 
\nonumber \\ 
& = & \frac{1}{2} \, \frac{1}{r_0 - \tau \overline{r} + i 
\epsilon (r_0) \gamma_q} \nonumber \\ 
& & + i \pi \epsilon (r_0) \, n_F (|r_0|) \, \displaystyle{ 
\raisebox{0.9ex}{\scriptsize{$\diamond$}}} \mbox{\hspace{-0.33ex}} 
\rho_\tau (R) \, , 
\label{saishuu-1} \\ 
\displaystyle{ \raisebox{1.1ex}{\scriptsize{$\diamond$}}} 
\mbox{\hspace{-0.33ex}} \tilde{S}_{1 2 / 2 1}^{(\tau)} (R) 
& \simeq & \pm i \pi n_F (\pm r_0) \, \displaystyle{ 
\raisebox{0.9ex}{\scriptsize{$\diamond$}}} \mbox{\hspace{-0.33ex}} 
\rho_\tau (R) \, , 
\label{saishuu-2} 
\end{eqnarray} 
where $\overline{r} \equiv r + m_f^2 / r$ and 
\begin{eqnarray} 
\raisebox{0.9ex}{\scriptsize{$\diamond$}} 
\mbox{\hspace{-0.33ex}} \rho_\tau (R) & = & \delta_{\gamma_q} [r_0 
- \tau \overline{r} ] \nonumber \\ 
& \equiv & \frac{1}{\pi} \, \frac{\gamma_q}{[r_0 - \tau \overline{r} 
]^2 + \gamma_q^2} \, . 
\label{sin-rho} 
\end{eqnarray} 
As mentioned above, the forms (\ref{saishuu-1}) - (\ref{sin-rho}) 
are gauge independent. 

Let us compare $\delta_{\gamma_q} (r_0 - \tau \overline{r})$ with 
$\delta (r_0 - \tau r)$, which is the bare counterpart of 
$\raisebox{0.9ex}{\scriptsize{$\diamond$}} \mbox{\hspace{-0.33ex}} 
\rho_\tau (R)$ (cf. Appendix A). $\delta (r_0 - \tau r)$ \lq\lq 
peaks'' at $r_0 = \tau r$, which shifts to $r_0 = \tau (r + m_f^2 / 
r)$ $=$ $\tau r + O (g^2 T)$ in $\delta_{\gamma_q} (r_0 - \tau 
\overline{r})$. The width of $\delta (r_0 - \tau r)$ is zero, while 
the width of $\delta_{\gamma_q} (r_0 - \tau \overline{r})$ is of $O 
(\gamma_q)$ $=$ $O (g^2 T \, \ln (g^{- 1}))$. Note that, at 
logarithmic accuracy, $\gamma_q$ $>>$ $m_f^2 / r$ $=$ $O (g^2 T)$. 
Similar observation can be made for $Re \, \displaystyle{ 
\raisebox{1.1ex}{\scriptsize{$\diamond$}}} \mbox{\hspace{-0.33ex}} 
\tilde{S}_{1 1}^{(\tau)} (R)$. Thus, the free hard-quark propagator 
is modified in the region 
\begin{equation} 
|r_0 - \tau r| \leq O (g^2 T \, \ln (g^{- 1})) \, , 
\label{app-region} 
\end{equation} 
with $\tau = \epsilon (r_0)$. More precisely \cite{sin-nie}, the 
forms of $Im \, \displaystyle{ 
\raisebox{1.1ex}{\scriptsize{$\diamond$}}} \mbox{\hspace{-0.33ex}} 
\tilde{S}_{1 1}^{(\tau)} (R)$ $(= Im \, \displaystyle{ 
\raisebox{1.1ex}{\scriptsize{$\diamond$}}} \mbox{\hspace{-0.33ex}} 
\tilde{S}_{2 2}^{(\tau)} (R))$ (cf. (\ref{saishuu-1})) and 
$\displaystyle{ \raisebox{1.1ex}{\scriptsize{$\diamond$}}} 
\mbox{\hspace{-0.33ex}} \tilde{S}_{1 2 / 2 1}^{(\tau)} (R)$ in 
(\ref{saishuu-2}) are valid in the region (\ref{app-region}), while 
the form of $Re \, \displaystyle{ 
\raisebox{1.1ex}{\scriptsize{$\diamond$}}} \mbox{\hspace{-0.33ex}} 
\tilde{S}_{1 1}^{(\tau)} (R)$ $(= - Re \, \displaystyle{ 
\raisebox{1.1ex}{\scriptsize{$\diamond$}}} \mbox{\hspace{-0.33ex}} 
\tilde{S}_{2 2}^{(\tau)} (R))$ is valid in the region $O [g^3 T]$ 
$<$ $|r_0 - \epsilon (r_0) \, \overline{r}|$ $\leq$ $O (g^2 T \, \ln 
(g^{- 1}))$. (For the definition of \lq\lq $O [g^3 T]$'', see the 
end of Sec. I.) 

Comparing (\ref{a-2}) and (\ref{a-3}) in Appendix A with 
(\ref{saishuu-1}) and (\ref{saishuu-2}), we see that, in (3.3b), 
the following substitutions should be made, 
\begin{eqnarray} 
\delta (q_0 - \tau q + K' \cdot \hat{Q}_\tau) & \to & 
\raisebox{0.9ex}{\scriptsize{$\diamond$}} \mbox{\hspace{-0.33ex}} 
\rho_\tau (Q + K') \simeq \frac{1}{\pi} \, \frac{\gamma_q}{[q_0 - 
\tau \overline{q} + K' \cdot \hat{Q}_\tau ]^2 + \gamma_q^2} \, , 
\nonumber \\ 
\frac{{\cal P}}{q_0 - \tau q + K \cdot \hat{Q}_\tau} & \to & 
\frac{{\cal P}_{\gamma_q}}{q_0 - \tau \overline{q} + K \cdot 
\hat{Q}_\tau} \equiv \frac{q_0 - \tau \overline{q} + K \cdot 
\hat{Q}_\tau }{[q_0 - \tau \overline{q} + K \cdot \hat{Q}_\tau ]^2 + 
\gamma_q^2} \, . 
\label{ver-20} 
\end{eqnarray} 

\noindent Let $\left( \raisebox{1ex}{\scriptsize{$\diamond$}} 
\mbox{\hspace{0.1ex}} \tilde{\Gamma}^\mu \right)^\ell_{i j}$ be the 
photon-quark vertex function that is obtained from 
$\left( \displaystyle{\raisebox{0.6ex}{\scriptsize{*}}} 
\tilde{\Gamma}^\mu  \right)^\ell_{i j}$, Eq. (\ref{ver-1}), through 
the above replacements, which is diagrammed in Fig. 4. The 
substitution (\ref{ver-20}) results in a violation of Ward-Takahashi 
relation, Eq. (\ref{ward}), on the basis of which our analysis is 
going. This issue will be dealt with in the following section. 
Substituting the formulas in Appendix A into (\ref{ver-1}) and 
making the replacements (\ref{ver-20}), we obtain for $\left( 
\raisebox{1ex}{\scriptsize{$\diamond$}} 
\mbox{\hspace{0.1ex}} \tilde{\Gamma}^\mu \right)^1_{2 1}$, 
\begin{eqnarray} 
& & \left( \raisebox{1ex}{\scriptsize{$\diamond$}} 
\mbox{\hspace{0.1ex}} \tilde{\Gamma}^\mu (K', K) \right)^1_{2 1} 
\nonumber \\ 
& & \mbox{\hspace*{5ex}} \simeq  2 \pi i g^2 \, C_F \int \frac{d^{\, 
4} Q}{ (2 \pi)^3} \sum_{\tau = \pm} \hat{Q}^\mu_\tau \, 
\hat{{Q\kern-0.1em\raise0.3ex\llap{/}\kern0.15em\relax}}_\tau \, 
n_F (- q_0) [\theta (- q_0) + n_B (q)] \, \delta (Q^2) \nonumber \\ 
& & \mbox{\hspace*{8ex}} \times 
\raisebox{0.9ex}{\scriptsize{$\diamond$}} \mbox{\hspace{-0.33ex}} 
\rho_\tau (Q + K') \, \frac{{\cal P}_{\gamma_q}}{q_0 - \tau 
\overline{q} + K \cdot \hat{Q}_\tau} \nonumber \\ 
& & 
\mbox{\hspace*{5ex}} 
\simeq \frac{i}{8 \pi} \, m_f^2 
\sum_{\tau = \pm} \int d \Omega \, \hat{Q}^\mu_\tau \, 
\hat{{Q\kern-0.1em\raise0.3ex\llap{/}\kern0.15em\relax}}_\tau \, 
\frac{\gamma_q}{(K' \cdot \hat{Q}_\tau - \tau m_f^2 / T)^2 + 
\gamma_q^2} \nonumber \\ 
& & 
\mbox{\hspace*{8ex}} 
\times \frac{K \cdot \hat{Q}_\tau - \tau m_f^2 / T}{(K \cdot 
\hat{Q}_\tau - \tau m_f^2 / T)^2 + \gamma_q^2} \, . 
\label{ver-21} \\ 
& & \mbox{\hspace*{5ex}} \simeq \frac{i}{16 \pi} \, m_f^2 \, Im \int 
d \Omega \, \hat{Q}^\mu \, 
\hat{{Q\kern-0.1em\raise0.3ex\llap{/}\kern0.15em\relax}} \nonumber 
\\ 
& & \mbox{\hspace*{8ex}} \times \sum_{\tau = \pm} \left[ \frac{1}{P 
\cdot \hat{Q}} \left( \frac{1}{\hat{Q} \cdot K' - \tau m_f^2 / T - i 
\gamma_q} - \frac{1}{\hat{Q} \cdot K - \tau m_f^2 / T - i \gamma_q} 
\right) \right. \nonumber \\ 
& & \mbox{\hspace*{8ex}} \left. - \frac{1}{P \cdot \hat{Q} - 2 i 
\gamma_q} \left( \frac{1}{\hat{Q} \cdot K' - \tau m_f^2 / T + i 
\gamma_q} - \frac{1}{\hat{Q} \cdot K - \tau m_f^2 / T - i \gamma_q} 
\right) \right] \, . 
\label{ver-211} 
\end{eqnarray} 
\noindent When integrating over $q = |{\bf q}|$ to obtain 
(\ref{ver-21}), the region $q = O (T)$ dominates, and then we have 
made the replacement: $\overline{q} = q + m_f^2 / q$ $\to$ $q + 
m_f^2 / T$. 

Singling out the contribution that diverges in the limit $\gamma_q$ 
$\to$ $0^+$, we have 
\begin{eqnarray} 
\left( \raisebox{1ex}{\scriptsize{$\diamond$}} 
\mbox{\hspace{0.1ex}} \tilde{\Gamma}^\mu (K', K) \right)^1_{2 1} 
& \simeq & \frac{i}{16} \, m_f^2 \int d \Omega \, \hat{Q}^\mu \, 
\hat{{Q\kern-0.1em\raise0.3ex\llap{/}\kern0.15em\relax}} \, 
\frac{P \cdot \hat{Q}}{(P \cdot \hat{Q})^2 + 4 \gamma_q^2} 
\nonumber \\ 
& & \times \sum_{\tau = \pm} \left\{ \delta_{\gamma_q} (\hat{Q} 
\cdot K' - \tau m_f^2 / T) \right. \nonumber \\ 
& & \left. + \delta_{\gamma_q} (\hat{Q} 
\cdot K - \tau m_f^2 / T) \right\} \, . 
\label{ver-36} 
\end{eqnarray} 
In the limit $\gamma_q , \, m_f^2 / T$ $\to$ $0^+$, $P \cdot \hat{Q} 
/ \{ (P \cdot \hat{Q})^2 + 4 \gamma_q^2 \}$ $\to$ $1 / P \cdot 
\hat{Q}$ diverges at ${\bf p} \parallel {\bf q}$, and the singular 
contribution (2.15a) is reproduced. 

Here a comment is in order. As mentioned at the beginning of this 
subsection, in Fig. 4, we should assign $\displaystyle{ 
\raisebox{1.1ex}{\scriptsize{$\diamond$}}} \mbox{\hspace{-0.33ex}} 
\Delta_{j i} (Q)$ to the hard-gluon propagator (cf. 
(\ref{shuppatsu})). Substitution of $\displaystyle{ 
\raisebox{1.1ex}{\scriptsize{$\diamond$}}} \mbox{\hspace{-0.33ex}} 
\Delta_{j i} (Q)$ for $\Delta_{j i} (Q)$ results in a change in 
$\delta_{\gamma_q}$s in (\ref{ver-36}). However, the features stated 
above after in conjunction with (\ref{app-region}) are unchanged, 
i.e., the point at which $\delta_{\gamma_q}$ peaks shifts by an 
amount of $O (g^2 T)$ and the width of $\delta_{\gamma_q}$ is of 
$O (g^2 T \, \ln (g^{- 1}))$. 

We may take the limit $\gamma_q, \, m_f^2 / T$ $\to$ $0^+$ in the 
quantity in the curly brackets in (\ref{ver-36}), since it does not 
yields any divergence at all (cf. observation in Sec. IIIA): 
\begin{eqnarray} 
\left( \raisebox{1ex}{\scriptsize{$\diamond$}} \mbox{\hspace{0.1ex}} 
\tilde{\Gamma}^\mu (K', K) \right)^1_{2 1} & \simeq & \frac{i}{8} \, 
\frac{m_f^2}{E} \int d \Omega \, \hat{Q}^\mu \, 
\hat{{Q\kern-0.1em\raise0.3ex\llap{/}\kern0.15em\relax}} \, 
\frac{1 - \hat{{\bf p}} \cdot \hat{{\bf q}}}{(1 - \hat{{\bf p}} 
\cdot \hat{{\bf q}})^2 + \tilde{\gamma}_q^2} \nonumber \\ 
& & \times \{ \delta (\hat{Q} \cdot K') + \delta (\hat{Q} \cdot K) 
\} \, , 
\label{3.20'} 
\end{eqnarray} 
where 
\begin{equation} 
\tilde{\gamma}_q \equiv 2 \gamma_q / E = O (g \, \ln (g^{- 1})) . 
\label{til-gamma} 
\end{equation} 
Thus, at logarithmic accuracy, we obtain 
\begin{eqnarray} 
\left( \raisebox{1ex}{\scriptsize{$\diamond$}} \mbox{\hspace{0.1ex}} 
\tilde{\Gamma}^\mu (K', K) \right)^1_{2 1} & \simeq & 
\frac{i \pi}{2} \, \frac{m_f^2}{E} \hat{P}^\mu \, 
\hat{{P\kern-0.1em\raise0.3ex\llap{/}\kern0.15em\relax}} \, 
\delta(\hat{P} \cdot K') \nonumber \\ 
& & \times \int d (\hat{{\bf p}} \cdot \hat{{\bf q}}) \, \frac{1 - 
\hat{{\bf p}} \cdot \hat{{\bf q}}}{(1 - \hat{{\bf p}} \cdot 
\hat{{\bf q}})^2 + \tilde{\gamma}_q^2} \nonumber \\ 
& \simeq & i \frac{\pi}{2} \, \frac{m_f^2}{E} \ln (1 / 
\tilde{\gamma}_q) \, \hat{P}^\mu \, 
\hat{{P\kern-0.1em\raise0.3ex\llap{/}\kern0.15em\relax}} \, \delta 
(\hat{P} \cdot K') \nonumber \\ 
& \simeq & i \frac{\pi}{2} \, \frac{m_f^2}{E} \ln (g^{- 1}) 
\hat{P}^\mu \, 
\hat{{P\kern-0.1em\raise0.3ex\llap{/}\kern0.15em\relax}} \, \delta 
(\hat{P} \cdot K') \, . \nonumber \\ 
\label{ver-362} 
\end{eqnarray} 
\noindent It should be noted that this contribution comes from the 
region, 
\begin{eqnarray} 
& & |q_i - |q_{i 0}|| = O (g^2 T \, \ln (g^{- 1})) \, , \;\;\;\;\;\; 
(i = 1, 2) \, , \nonumber \\ 
& & O (g \, \ln (g^{- 1})) \leq 1 - \hat{{\bf p}} \cdot 
\hat{{\bf q}} << 1 \, , 
\label{area} 
\end{eqnarray} 
where $Q_1 \equiv Q + K'$ and $Q_2 \equiv Q + K$ (cf. 
(\ref{app-region}), (\ref{ver-1}) and (\ref{ver-21})). 

Through similar analysis, we obtain 
\begin{equation} 
\left( \raisebox{1ex}{\scriptsize{$\diamond$}} \mbox{\hspace{0.1ex}} 
\tilde{\Gamma}^\mu (K', K) \right)^1_{1 2} \simeq i \frac{\pi}{2} \, 
\frac{m_f^2}{E} \, \ln (g^{- 1}) \, \hat{P}^\mu \, 
\hat{{P\kern-0.1em\raise0.3ex\llap{/}\kern0.15em\relax}} \, \delta 
(K \cdot \hat{P}) \, . 
\label{ver-38} 
\end{equation} 
\subsection{Contribution to the soft-photon production rate} 
Comparing (\ref{ver-362}) and (\ref{ver-38}) with 
(\ref{HTL-vertex}), we see that the replacements (\ref{ver-20}) make 
the singular contribution (\ref{w2}) the finite contribution, 
\begin{eqnarray} 
E \, \frac{dW}{d^{\, 3} p} 
& = & \frac{e_q^2 e^2 N_c}{(2 \pi)^3} \, \frac{1}{E} \, Re \int 
\frac{d^{\, 4} K}{(2 \pi)^4} \, tr \left[ \displaystyle{ 
\raisebox{0.6ex}{\scriptsize{*}}} \! S_{2 i} (K') \left( 
\displaystyle{\raisebox{1ex}{\scriptsize{$\diamond$}}} 
\mbox{\hspace{0.1ex}} \tilde{\Gamma}^0 (K', K) \right)^1_{i 2} 
\right. \nonumber \\ 
& & \left. - \left( \displaystyle{ 
\raisebox{1ex}{\scriptsize{$\diamond$}}} \mbox{\hspace{0.1ex}} 
\tilde{\Gamma}^0 (K', K) \right)^1_{2 i} \displaystyle{ 
\raisebox{0.6ex}{\scriptsize{*}}} \! S_{i 2} (K) \right] \, , 
\nonumber \\ 
& \simeq & \frac{e_q^2 \, \alpha \, \alpha_s}{2 \pi^2} \, T^2 \left( 
\frac{m_f}{E} \right)^2 \, \ln (g^{- 1}) \ln \left( \frac{k^*}{m_f} 
\right) \, , 
\label{yuugenn} 
\end{eqnarray} 
\noindent which is valid at logarithmic accuracy. As has been 
already mentioned, the result (\ref{yuugenn}) is gauge independent. 

As to the hard contribution, the mass singularity is cutoff 
\cite{reb1} in the same way as in the soft contribution analyzed 
above, which produces a $\ln (g^{- 1})$ as in (\ref{yuugenn}). 
Adding the hard contribution to (\ref{yuugenn}), we obtain 
\begin{eqnarray} 
E \, \frac{dW}{d^{\, 3} p} & \simeq & \frac{e_q^2 \, \alpha \, 
\alpha_s}{2 \pi^2} \, T^2 \left( \frac{m_f}{E} \right)^2 \, \ln 
(g^{- 1}) \ln \left( \frac{T}{m_f} \right) \nonumber \\ 
& \simeq & \frac{e_q^2 \, \alpha \, \alpha_s}{2 \pi^2} \, T^2 
\left( \frac{m_f}{E} \right)^2 \, \ln^2 (g^{- 1}) \, . 
\label{katai-1} 
\end{eqnarray} 

In computing $d W / d^{\, 3} p$, Eq. (\ref{yuugenn}), for soft-quark 
propagators, we have used $\displaystyle{ 
\raisebox{0.6ex}{\scriptsize{*}}} \! S$s, which is evaluated in 
canonical HTL-resummation scheme. However, to be consistent, in 
computing the soft-quark self-energy part, Fig. 5, $\displaystyle{ 
\raisebox{1.1ex}{\scriptsize{$\diamond$}}} \mbox{\hspace{-0.33ex}} 
S$ and $\displaystyle{ \raisebox{1.1ex}{\scriptsize{$\diamond$}}} 
\mbox{\hspace{-0.33ex}} \Delta$ should be assigned, in respective 
order, to the quark- and gluon-lines in the HTL. $\displaystyle{ 
\raisebox{0.6ex}{\scriptsize{*}}} \! S (K)$ is written in terms of 
$D_\sigma (K)$ $(\sigma = \pm)$, Eq. (\ref{eff5}) in Appendix A. Let 
us briefly see how does the form of $D_\sigma (K)$ change by the 
above-mentioned replacements, $S$ $\to$ $\displaystyle{ 
\raisebox{1.1ex}{\scriptsize{$\diamond$}}} \mbox{\hspace{-0.33ex}} 
S$ and $\Delta$ $\to$ $\displaystyle{ 
\raisebox{1.1ex}{\scriptsize{$\diamond$}}} \mbox{\hspace{-0.33ex}} 
\Delta$. In $D_\sigma (K)$, the term $\sigma m_f^2 / k$ is 
insensitive to the region $Q^2 \simeq 0$ and/or $(K - Q)^2 \simeq 0$ 
in Fig. 5. Then, the change in the term $\sigma m_f^2 / k$ is of 
higher order. The logarithmic factor in $D_\sigma (K)$, $\ln [\{ k_0 
(1 + 0^+) + k \}/\{ k_0 (1 + 0^+) - k \}]$, changes to 
\[ 
\ln \left( \frac{k_0 + k + a + i \epsilon (k_0) \, b}{k_0 - k + a + 
i \epsilon (k_0) \, b} \right) \, , 
\] 
where $|a| = O (g^2 T)$ and $b = O (g^2 T \, \ln (g^{- 1}))$. 
Therefore, the change in the logarithmic factor is appreciable only 
in the region, $|k - |k_0||$ $\leq$ $O [g^2 T]$, which is small when 
compared to the whole soft $(k_0, k)$-region ($|k_0| \leq k$). Then 
the leading-order contribution (\ref{yuugenn}) is not affected by 
this change (cf. (\ref{bare-final}) and \cite{bai1}). 
\setcounter{equation}{0}
\setcounter{section}{3}
\section{Photon-quark vertex corrections} 
\def\theequation{\mbox{\arabic{section}.\arabic{equation}}}
\subsection{Preliminary} 
Deducing the result (\ref{katai-1}) is not the end of the analysis. 
In deriving (\ref{pil22}) or (\ref{yuugenn}), we have used the 
Ward-Takahashi relation (\ref{ward}), which is a representation of 
the current-conservation condition or the gauge invariance. 
Substitution of $\raisebox{1ex}{\scriptsize{$\diamond$}} 
\mbox{\hspace{0.1ex}} \tilde{\Gamma}^\mu$ (cf. (\ref{ver-21})) for 
$\displaystyle{\raisebox{0.6ex}{\scriptsize{*}}} \tilde{\Gamma}^\mu$ 
in (\ref{ward}) violates the current-conservation condition. 

For recovering it, one needs to include corrections to the 
photon-quark vertex in Fig. 4. Here we face the 
question: In the kinematical region of our interest, Eq. 
(\ref{area}), what kind of diagrams does participate. In other 
wards, what kind of vertex corrections leads to the contribution, 
which is of the same order of magnitude as the bare photon-quark 
vertex. 

Similar problem arises in the damping rate of a moving 
(quasi)particle in a hot QCD/QED plasma. Lebedev and Smilga 
\cite{l-s} have shown that the relevant diagrams are the ladder 
diagrams as depicted in Fig. 6, where all the gluon rungs carry the 
soft momenta. (In Fig. 6, solid- and dashed-lines stand, 
respectively, for quark- and gluon-propagators, $P$ is soft, and 
$Q_1$ (and then also $Q_2$ ($=$ $Q_1 + P$) is hard). The region of 
our interest is (cf. (\ref{area})) 
\begin{eqnarray} 
& & |q_{j 0} - \epsilon (q_{j 0}) q_j| = O (\gamma_q) = O (g^2 T \, 
\ln (g^{- 1})) \, , \nonumber \\ 
& & \mbox{\hspace*{34ex}} (j = 1, 2) \, , 
\label{kawa-1} \\ 
& & E - \tau \hat{{\bf q}}_1 \cdot {\bf p} = O (\gamma_q) \, , 
\label{kawa-11} 
\end{eqnarray} 
where $\gamma_q$ is as in (\ref{yama-2}). {\em At logarithmic 
accuracy}, the leading contribution comes \cite{l-s} from the 
magnetic sector of the effective gluon propagators. That some other 
diagrams than Fig. 6 lead to nonleading contributions are discussed 
in \cite{l-s,sin-nie}. 

The contribution from Fig. 6 with $n$-rungs reads \cite{l-s} 
\begin{equation} 
\left( \hat{\Lambda}^\mu_n (Q_1, Q_2) \right)^\ell_{j i} \simeq - 
(-)^{i + j + \ell} \gamma^0\, \hat{Q}_{1 \tau}^\mu \sum_{\rho = \pm} 
\left[ {\cal N}^{(\rho)}_{j \ell} {\cal N}^{(- \rho)}_{\ell i} 
\left\{ \frac{- 2 i \rho \tau \gamma_q}{E - \tau \hat{{\bf q}}_1 
\cdot {\bf p} - 2 i \rho \tau \gamma_q} \right\}^n \right] \, , 
\label{ls-01} 
\end{equation} 
\noindent where the sum is not taken over $\ell$, $j$, and $i$, $Q_2 
- Q_1$ $=$ $P$, and 
\begin{eqnarray} 
{\cal N}^{(+)}_{1 1} & = & - {\cal N}^{(-)}_{2 2} = 1 - n_F (q_1) 
\nonumber \\ 
{\cal N}^{(-)}_{1 1} & = & - {\cal N}^{(+)}_{2 2} = n_F (q_1) 
\nonumber \\ 
{\cal N}^{(+)}_{1 2 / 2 1} & = & - {\cal N}^{(-)}_{1 2 / 2 1} = 
\theta (\mp q_{1 0}) - n_F (q_1) \, . 
\label{ls-03} 
\end{eqnarray} 

$\hat{\Lambda}^\mu_n$ in (\ref{ls-01}) meets (\ref{recip}), which 
serves as a cross-check of the validity of (\ref{ls-01}). 

It is straightforward to resum $\hat{\Lambda}^\mu_n (Q_1, Q_2)$ over 
$n$: 
\begin{mathletters} 
\label{karage-1} 
\begin{eqnarray} 
& & \left( \hat{\Lambda}^\mu (Q_1, Q_2) \right)^\ell_{j i} \equiv 
\sum_{n = 1}^\infty \left[ \hat{\Lambda}^\mu_n (Q_1, Q_2) 
\right]^\ell_{j i} \, , \nonumber \\ 
& & \mbox{\hspace*{34ex}} (\ell = 1, 2) \, , 
\eqnum{4.5a} 
\label{ls-0} \\ 
& & \left( \hat{\Lambda}^\mu (Q_1, Q_2) \right)^\ell_{1 1} \simeq 
\left( \hat{\Lambda}^\mu (Q_1, Q_2) \right)^\ell_{2 2} \simeq 0 \, , 
\eqnum{4.5b} 
\label{3.3'} \\ 
& & \left( \hat{\Lambda}^\mu (Q_1, Q_2) \right)^\ell_{1 2 / 2 1} 
\nonumber \\ 
& & \mbox{\hspace*{4ex}} \simeq \pm 2 i \, \tau (-)^\ell \, 
\gamma^0 \, \hat{Q}_{1 \tau}^\mu [ \theta (\mp q_{1 0}) - n_F (q_1) 
] \frac{\gamma_q}{E - \tau \hat{{\bf q}}_1 \cdot {\bf p}} \, , 
\nonumber \\ 
\eqnum{4.5c} 
\label{ls-1} 
\end{eqnarray} 
\end{mathletters} 

\noindent where $\tau = \epsilon (q_{1 0})$. Thus, through 
resummation, the imaginary part of the denominator in 
$\hat{\Lambda}^\mu_{n = 1} (Q_1, Q_2)$, Eq. (\ref{ls-01}), 
\lq\lq disappears'', which is the important finding in \cite{l-s}. 
[Kraemmer, Rebhan, and Schultz \cite{k-r-s} have discussed that the 
same phenomenon takes place in scalar QCD.] 

It should be emphasized that the result (\ref{ls-1}) is valid to 
leading order {\em at logarithmic accuracy}, i.e., valid in the 
region, (\ref{kawa-1}) and (\ref{kawa-11}). More precisely, 
$\gamma_q$ in the numerator of (\ref{ls-01}) and $\gamma_q$ in the 
denominator is the same only at logarithmic accuracy, $\gamma_q$ 
$=$ $O ( g^2 T \ln (g^{- 1}) )$. 

It is straightforward to show \cite{l-s} that, to the accuracy we 
are taking, the photon-quark vertex function 
\begin{equation} 
\left( \Lambda^\mu (Q_1, Q_2) \right)^1_{j i} \equiv \delta_{1 i} \, 
\delta_{1 j} \, \gamma^\mu + \left( \hat{\Lambda}^\mu (Q_1, Q_2) 
\right)^1_{j i} 
\label{add-t} 
\end{equation} 
satisfies the Ward-Takahashi relation, 
\begin{eqnarray} 
P_\mu \left( \Lambda^\mu (Q_1, Q_2) \right)^1_{j i} 
& \simeq & \delta_{1 j} \displaystyle{ 
\raisebox{1.1ex}{\scriptsize{$\diamond$}}} \mbox{\hspace{-0.33ex}} 
S^{- 1}_{1 i} ( Q_2)  - \delta_{1 i} \displaystyle{ 
\raisebox{1.1ex}{\scriptsize{$\diamond$}}} \mbox{\hspace{-0.33ex}} 
S^{- 1}_{j 1} ( Q_1) \, , 
\label{WT-ne} 
\end{eqnarray} 
(with $P = Q_2 - Q_1$). $\displaystyle{ 
\raisebox{1.1ex}{\scriptsize{$\diamond$}}} \mbox{\hspace{-0.33ex}} 
S^{- 1}_{j i} (Q)$ is the inverse-matrix function of $\displaystyle{ 
\raisebox{1.1ex}{\scriptsize{$\diamond$}}} \mbox{\hspace{-0.33ex}} 
S_{j i} (Q)$: 
\begin{mathletters} 
\label{karage-2}
\begin{eqnarray} 
\displaystyle{ \raisebox{1.1ex}{\scriptsize{$\diamond$}}} 
\mbox{\hspace{-0.33ex}} S^{- 1}_{1 1} (Q) & = & - 
\left[ \displaystyle{ \raisebox{1.1ex}{\scriptsize{$\diamond$}}} 
\mbox{\hspace{-0.33ex}} S^{- 1}_{2 2} (Q) \right]^* \nonumber \\ 
& = & {Q\kern-0.1em\raise0.3ex\llap{/}\kern0.15em\relax} 
- \tilde{\Sigma}_F (Q) + 2 i n_F (q) \, Im \tilde{\Sigma}_F (Q) 
\eqnum{4.8a} 
\label{ora} \\ 
& \simeq & {Q\kern-0.1em\raise0.3ex\llap{/}\kern0.15em\relax} - \tau 
\, \gamma^0 [ m_f^2/ q - i \{ 1 - 2 n_F (q) \} \, \gamma_q ] \, , 
\nonumber \\ 
\eqnum{4.8b} 
\label{sub} \\ 
\displaystyle{ \raisebox{1.1ex}{\scriptsize{$\diamond$}}} 
\mbox{\hspace{-0.33ex}} S^{- 1}_{1 2 / 2 1} (Q) & = & 2 i [ \theta 
(\mp q_0) - n_F (q) ] \, Im \tilde{\Sigma}_F (Q) 
\eqnum{4.8c} 
\label{ora-1} \\ 
& \simeq & - 2 i \tau \, \gamma^0 [ \theta (\mp q_0) - n_F (q) ] \, 
\gamma_q \, . 
\eqnum{4.8d} 
\label{add} 
\end{eqnarray} 
\end{mathletters} 

\noindent As a matter of fact, using (\ref{karage-1}) and 
(\ref{karage-2}), we see that the difference between the L.H.S. of 
(\ref{WT-ne}) and the R.H.S. is of $O [g^3 T]$. 
\subsection{Estimate of the form for $\hat{\Lambda}^\mu (Q_1, Q_2)$} 
For the purpose of inferring the form of $\hat{\Lambda}^\mu (Q_1, 
Q_2)$ that holds in much wider region than (\ref{kawa-11}), 
we reverse the order of argument. Namely, after {\em imposing} 
Ward-Takahashi relation (\ref{WT-ne}), diagrammatic analysis 
follows. The region of our interest here is 
\begin{equation} 
\Delta \, T \equiv |E - \tau \hat{{\bf q}}_1 \cdot {\bf p}| 
\leq O (g^2 T \, \ln (g^{-1 })) \, . 
\label{hiroi} 
\end{equation} 
It is sufficient to analyze $\left( \hat{\Lambda}^\mu \right)^1_{i 
j}$, since $\left( \hat{\Lambda}^\mu \right)^2_{i j}$ is obtained 
from $\left( \hat{\Lambda}^\mu \right)^1_{i j}$ through 
(\ref{recip}). 

We first make preliminary remarks. Since $|\tilde{\Sigma}_F (Q_1)| = 
O [g^2 T]$ and $|\tilde{\Sigma}_F (Q_2) - \tilde{\Sigma}_F (Q_1)|$ 
$=$ $O [g^3 T]$, (\ref{WT-ne}) with (\ref{karage-2}) yields 
\begin{eqnarray} 
& & \rule[-3mm]{.14mm}{8.5mm} \, P_\mu \left( \hat{\Lambda}^\mu 
\right)^1_{11 / 22} \rule[-3mm]{.14mm}{8.5mm} = O [g^3 T] \, , 
\;\;\;\;\;\; \rule[-3mm]{.14mm}{8.5mm} \, P_\mu \left( 
\hat{\Lambda}^\mu \right)^1_{12 / 21} \rule[-3mm]{.14mm}{8.5mm} 
= O [g^2 T] \, . \nonumber \\ 
\label{moku-1} 
\end{eqnarray} 
This fact, together with (\ref{3.3'}), indicates that, for our 
purpose, $\left( \hat{\Lambda}^\mu \right)^1_{11 / 22}$ may be 
ignored. Then we concentrate our concerns on $\left( 
\hat{\Lambda}^\mu \right)^1_{12 / 21}$. Eq. (\ref{moku-1}) together 
with $|P^\mu|$ $=$ $O (g T)$ also shows that it is sufficient to 
analyze the structure of $\left( \hat{\Lambda}^\mu \right)^1_{1 2 / 
2 1}$ up to and including $O [g]$ contribution, with the proviso to 
be discussed below. Furthermore, as in Secs. II and III and as will 
be shown below, we need only $\left( \hat{\Lambda}^{\mu = 0} 
\right)^1_{12 / 21}$ {\em to leading order}, provided that $\left( 
\hat{\Lambda}^\mu \right)^1_{12 / 21}$ satisfies the Ward-Takahashi 
relation. Whenever confusion does not arise in the following, we 
drop thermal indices. The last remark is on the $4 \times 4$ matrix 
structure of $\hat{\Lambda}^\mu (Q_1, Q_2)$. We recall that in 
computing the modified HTL contribution to the photon-quark vertex, 
$\hat{{Q\kern-0.1em\raise0.3ex\llap{/}\kern0.15em\relax}}_{1 \tau}$ 
($\hat{{Q\kern-0.1em\raise0.3ex\llap{/}\kern0.15em\relax}}_{2 
\tau}$) is to be multiplied from the left (right) of 
$\hat{\Lambda}^{\mu = 0} (Q_1, Q_2)$. Then the term in 
$\hat{\Lambda}^{\mu = 0} (Q_1, Q_2)$ that is proportional to 
$\hat{{Q\kern-0.1em\raise0.3ex\llap{/}\kern0.15em\relax}}_{1 \tau}$ 
$\simeq$ 
$\hat{{Q\kern-0.1em\raise0.3ex\llap{/}\kern0.15em\relax}}_{2 \tau}$ 
may be ignored. This is because 
$\left( \hat{{Q\kern-0.1em\raise0.3ex\llap{/}\kern0.15em\relax}}_{1 
\tau} \right)^2 = 0$ and, from (\ref{hiroi}), 
$|\hat{{Q\kern-0.1em\raise0.3ex\llap{/}\kern0.15em\relax}}_{1 \tau} 
\, \hat{{Q\kern-0.1em\raise0.3ex\llap{/}\kern0.15em\relax}}_{2 
\tau}|$ is at most of $O [g^{3 / 2}]$. 
\subsubsection{Resummation of ladder diagrams} 
With the above preliminary remarks in mind, we start with the 
analysis of (resummation of) ladder diagrams, Fig. 6, which yields 
\cite{l-s} the leading contribution. The region, where 
momenta of the quark lines adjacent to each gluon rung are soft, 
$| Q^{(j) \, \mu} | = O (g T) = | R^{(j) \, \mu} |$, is unimportant, 
because the phase-space volume is small. The contribution from the 
region, $\rule[-2mm]{.14mm}{6.5mm} \, \left( Q^{(j)} \right)^2 
\rule[-2mm]{.14mm}{6.5mm} = O (T^2) = \rule[-2mm]{.14mm}{6.5mm} \, 
\left( R^{(j)} \right)^2 \rule[-2mm]{.14mm}{6.5mm} \,$, is of $O \{ 
g^2 \}$. (\lq\lq $O \{ g^2 \}$'' is defined at the end of Sec. I.) 
The leading contribution comes from the region 
$\rule[-2mm]{.14mm}{6.5mm} \, \left( R^{(j)} \right)^2 
\rule[-2mm]{.14mm}{6.5mm} \, $, $\rule[-2mm]{.14mm}{6.5mm} \, 
\left( Q^{(j)} \right)^2 \rule[-2mm]{.14mm}{6.5mm} \,$ $\leq$ $O 
(\gamma_q T)$ $=$ $O [g^2 T]$ $(j = 2, ..., n + 1)$. Then it is 
sufficient to assume that each $K_j$ ($j = 1, ..., n$) is either 
soft or hard with nearly (anti)collinear to $Q_1^\mu$ $(\simeq 
Q_2^\mu)$. 

In Appendix B, we show that the contribution of Fig. 6, where at 
least one $K_j$ out of $K_2 - K_n$ is hard, is of $O [g^2]$. The 
contribution of Fig. 6 with all $K$s but $K_1$ are soft is of the 
form 
\begin{eqnarray*} 
\hat{\Lambda}^\mu (Q_1, Q_2) & = & {\cal F}_1^\mu (Q_1, Q_2) \, 
\hat{{Q\kern-0.1em\raise0.3ex\llap{/}\kern0.15em\relax}}_{1 \tau} + 
{\cal F}_2^\mu (Q_1, Q_2) \, 
\hat{{Q\kern-0.1em\raise0.3ex\llap{/}\kern0.15em\relax}}_{2 \tau} \\ 
& & + O \{ g^2 \} \, . 
\end{eqnarray*} 
Thus, according to the preliminary remarks above, we can ignore this 
contribution. 

Let us turn to analyze Fig. 6 with all $K$s soft. We pick out the 
term 
\[ 
\hat{{Q\kern-0.1em\raise0.3ex\llap{/}\kern0.15em\relax}}_\tau \, 
\gamma^\mu \, 
\hat{{R\kern-0.1em\raise0.3ex\llap{/}\kern0.15em\relax}}_\tau \, 
\left( \equiv 
\hat{{Q\kern-0.1em\raise0.3ex\llap{/}\kern0.15em\relax}}_\tau^{(n 
+ 1)} \, \gamma^\mu \, 
\hat{{R\kern-0.1em\raise0.3ex\llap{/}\kern0.15em\relax}}_\tau^{(n + 
1)} \right) \, , 
\] 
with $\tau = \epsilon (q_{1 0}) = \epsilon (q_{2 0})$. Noting that 
$\left( 
\hat{{Q\kern-0.1em\raise0.3ex\llap{/}\kern0.15em\relax}}_\tau 
\right)^2 = 0$, we obtain  
\begin{eqnarray} 
\hat{{Q\kern-0.1em\raise0.3ex\llap{/}\kern0.15em\relax}}_\tau \, 
\gamma^\mu \, 
\hat{{R\kern-0.1em\raise0.3ex\llap{/}\kern0.15em\relax}}_\tau 
& = & 2 \hat{Q}_\tau^\mu \, 
\hat{{R\kern-0.1em\raise0.3ex\llap{/}\kern0.15em\relax}}_\tau 
+ \tau \, \gamma^\mu \, 
\hat{{Q\kern-0.1em\raise0.3ex\llap{/}\kern0.15em\relax}}_\tau \, 
\frac{1}{q} \left[ \vec{\gamma} \cdot \{{\bf p} - ({\bf p} \cdot 
\hat{{\bf q}}) \, \hat{{\bf q}}\} \right] \, . \nonumber \\ 
\label{sindo-1} 
\end{eqnarray} 
\begin{center} 
{\em The first term on the R.H.S. of (\ref{sindo-1})}
\end{center} 

We first study the contribution, $\displaystyle{ 
\raisebox{1.1ex}{\scriptsize{$(1)$}}} \mbox{\hspace{-0.3ex}} 
\hat{\Lambda}^\mu (Q_1, Q_2)$, arising from the first term on the 
R.H.S. For $\displaystyle{ \raisebox{1.1ex}{\scriptsize{$(1)$}}} 
\mbox{\hspace{-0.3ex}} \hat{\Lambda}^{\mu = 0} (Q_1, Q_2)$, the 
first term reads $2 \hat{Q}_\tau^{\mu = 0} \, 
\hat{{R\kern-0.1em\raise0.3ex\llap{/}\kern0.15em\relax}}_\tau = 2 
\hat{{R\kern-0.1em\raise0.3ex\llap{/}\kern0.15em\relax}}_\tau$. 

For studying $P_\mu \displaystyle{ 
\raisebox{1.1ex}{\scriptsize{$(1)$}}} \mbox{\hspace{-0.3ex}} 
\hat{\Lambda}^\mu$, we begin with 
\begin{eqnarray} 
P \cdot \hat{Q}_\tau & = & P \cdot \hat{Q}_{1 \tau} - 
\frac{\tau}{q_1} ({\bf p} \cdot \underline{{\bf k}}_\perp) 
\nonumber \\ 
& & + \frac{\tau}{2 q_1^2} \left[ 2 (\hat{{\bf q}}_1 \cdot 
\underline{{\bf k}}) ({\bf p} \cdot \underline{{\bf k}}_\perp) + 
(\underline{{\bf k}}_\perp)^2 ({\bf p} \cdot \hat{{\bf q}}_1) 
\right] \nonumber \\ 
& & + O (g^4) \, , 
\label{dame-1} 
\end{eqnarray} 
where $\underline{{\bf k}}_\perp \equiv \underline{{\bf k}} - 
(\underline{{\bf k}} \cdot \hat{{\bf q}}_1) \, \hat{{\bf q}}_1$ 
with 
\begin{equation} 
\underline{{\bf k}} \equiv \sum_{j = 1}^n {\bf k}_j \, , 
\label{sum-yo} 
\end{equation} 
which is soft. The last term on the R.H.S. of (\ref{dame-1}) is of 
$O (g^3 T)$ and, according to the preliminary remarks above, it 
seems to be ignored. However, (\ref{hiroi}) tells us that the first 
term on the R.H.S. is of $O [\Delta T]$ with $\Delta \leq O [g^2]$. 
Then, for the purpose of finding the limit (of $\Delta$) of validity 
of the form $\hat{\Lambda}^\mu$ obtained below, we keep the last 
term in question. 

From the second term on the R.H.S. of (\ref{dame-1}), we take up 
${\bf p} \cdot {\bf k}_\perp / q_1 = p \, k \, \hat{{\bf p}} \cdot 
\hat{{\bf k}}_\perp / q_1$ $(K^\mu \equiv K_n^\mu)$ and trace 
$\hat{{\bf p}} \cdot \hat{{\bf k}}_\perp$, which appears in the 
integral 
\begin{equation} 
\langle 
\hat{{\bf p}} \cdot \hat{{\bf k}}_\perp \rangle \equiv \int d 
\Omega_{\hat{{\bf k}}} \, \hat{{\bf p}} \cdot \hat{{\bf k}}_\perp 
\, \displaystyle{ \raisebox{1.1ex}{\scriptsize{$\diamond$}}} 
\mbox{\hspace{-0.33ex}} \tilde{S}^{(\tau)} (Q) \, 
\displaystyle{ \raisebox{1.1ex}{\scriptsize{$\diamond$}}} 
\mbox{\hspace{-0.33ex}} \tilde{S}^{(\tau)} (R) \, . 
\label{gi-0} 
\end{equation} 
Here $d \Omega_{\hat{{\bf k}}}$ stands for the integration over the 
direction of $\hat{{\bf k}}$. $\displaystyle{ 
\raisebox{1.1ex}{\scriptsize{$\diamond$}}} \mbox{\hspace{-0.33ex}} 
\tilde{S}^{(\tau)} (Q)$ may be written as (cf. (\ref{saishuu-1}) - 
(\ref{sin-rho})) 
\begin{eqnarray} 
\displaystyle{ \raisebox{1.1ex}{\scriptsize{$\diamond$}}} 
\mbox{\hspace{-0.33ex}} \tilde{S}^{(\tau)} (Q) & = & \frac{1}{2} 
\sum_{\rho = \pm} \frac{{\cal N}^{(\rho)} (Q)}{q_0 - \tau 
\overline{q} + i \rho \tau \Gamma (q)} 
\label{pre-425} \\ 
& \simeq & \frac{1}{2} \sum_{\rho = \pm} \frac{{\cal N}^{(\rho)} 
(Q^{(n)})}{q_0^{(n)} - \tau \overline{q}^{(n)} + K \cdot 
\hat{Q}_\tau^{(n)} + i \rho \tau \Gamma (q^{(n)})} \, , \nonumber \\ 
\label{gi-1} 
\end{eqnarray} 
where ${\cal N}^{(\rho)}$s are as in (\ref{ls-03}) and $\Gamma (q)$ 
$=$ $\gamma_q$ $+$ $O (g^2 T)$ with $\gamma_q$ as in (\ref{yama-2}). 
Similarly, 
\begin{eqnarray} 
\displaystyle{ \raisebox{1.1ex}{\scriptsize{$\diamond$}}} 
\mbox{\hspace{-0.33ex}} \tilde{S}^{(\tau)} (R) 
& \simeq & \frac{1}{2} \sum_{\sigma = \pm} \frac{{\cal N}^{(\sigma)} 
(R)}{r_0^{(n)} - \tau \overline{r}^{(n)} + K \cdot 
\hat{R}_\tau^{(n)} + i \sigma \tau \Gamma (r)} \nonumber \\ 
& \simeq & \frac{1}{2} \sum_{\sigma = \pm} \frac{{\cal N}^{(\sigma)} 
(Q^{(n)})}{r_0^{(n)} - \tau \overline{r}^{(n)} + K \cdot 
\hat{Q}_\tau^{(n)} - \tau {\bf k} \cdot {\bf p}_T / q^{(n)} + i 
\sigma \tau \Gamma (q^{(n)})} \, , 
\label{gi-2} 
\end{eqnarray} 

\noindent where ${\bf p}_T \equiv {\bf p} - ({\bf p} \cdot 
\hat{{\bf q}}^{(n)}) \, \hat{{\bf q}}^{(n)}$. The dominant 
contribution comes from where the denominators of (\ref{gi-1}) and 
(\ref{gi-2}) are of $O [g^2 T]$. To the approximation we are keeping 
in mind, 
\[ 
\hat{{\bf k}}_\perp \simeq \hat{{\bf k}}_T = \hat{{\bf k}} - 
(\hat{{\bf k}} \cdot \hat{{\bf q}}^{(n)}) \, \hat{{\bf q}}^{(n)} 
\, . 
\] 
We take the direction $\hat{{\bf q}}^{(n)}$ as the $z$-axis and the 
direction ${\bf p}_T$ as the $x$-axis. Then $\hat{{\bf p}} \cdot 
\hat{{\bf k}}_\perp$ $\simeq$ $\hat{{\bf p}} \cdot \hat{{\bf k}}_T$ 
$=$ $O [(\Delta / g)^{1 / 2}] \cos \phi$ with $\phi$ the azimuth. 
$\displaystyle{ \raisebox{1.1ex}{\scriptsize{$\diamond$}}} 
\mbox{\hspace{-0.33ex}} \tilde{S}^{(\tau)} (Q)$ in (\ref{gi-1}) is 
independent of $\phi$. In $\displaystyle{ 
\raisebox{1.1ex}{\scriptsize{$\diamond$}}} \mbox{\hspace{-0.33ex}} 
\tilde{S}^{(\tau)} (R)$, only term that depends on $\phi$ is $ - 
\tau {\bf k} \cdot {\bf p}_T / q^{(n)}$: 
\begin{eqnarray*} 
- \tau \, \frac{{\bf k} \cdot {\bf p}_T}{q^{(n)}} & = & - \tau 
\frac{k_T \, p_T}{q^{(n)}} \, \cos \phi \\ 
& = & O [g^{3 / 2} \Delta^{1 / 2} T] \leq O [g^{5 / 2} T] \, . 
\end{eqnarray*} 
Then, the relative order of magnitude of $- \tau \, {\bf k} 
\cdot {\bf p}_T / q^{(n)}$ in the denominator of (\ref{gi-2}) is of 
$O [g^{3 / 2} \Delta^{1 / 2} T] / O [g^2 T]$ $=$ $O [(\Delta / g)^{1 
/ 2}]$. 

After all this, we find 
\[ 
\frac{\langle \hat{{\bf p}} \cdot \hat{{\bf k}}_\perp 
\rangle}{\langle 1 \rangle} = O [\Delta / g] \, , 
\] 
or 
\begin{equation} 
\frac{\langle {\bf p} \cdot {\bf k}_\perp / q_1\rangle}{\langle 1 
\rangle} = O [g \Delta T] \, , 
\label{tri-1} 
\end{equation} 
where $\langle 1 \rangle$ is defined by (\ref{gi-0}) with 
$\hat{{\bf p}} \cdot \hat{{\bf k}}_\perp$ deleted. It is not 
difficult to see that undoing the approximation, $\hat{{\bf 
k}}_\perp \simeq \hat{{\bf k}}_T$, used above, affects the term of 
$O [g^2 \Delta T]$ in (\ref{tri-1}). 

Similar analysis goes for $\hat{{\bf p}} \cdot \hat{{\bf 
k}}_{j \perp}$ $(1 \leq j \leq n - 1)$ in the second term on the 
R.H.S. of (\ref{dame-1}) and the same conclusion as above results. 

From the above analysis, we see that $P \cdot \hat{Q}_\tau$ in 
(\ref{dame-1}) turns out to 
\begin{equation} 
P \cdot \hat{Q}_\tau = E - \tau {\bf p} \cdot \hat{{\bf q}}_1 \left[ 
1 - \frac{1}{2 q_1^2} \, (\underline{{\bf k}}_\perp)^2 \right] + 
\lq\lq O\mbox{''} [g \Delta T] \, , 
\label{yare-1} 
\end{equation} 
where $\lq\lq O\mbox{''} [g \Delta T]$  means the term that leads to 
$O \{ g \Delta T \}$ contribution to $P_\mu \hat{\Lambda}^\mu$. We 
note that $E - \tau {\bf p} \cdot \hat{{\bf q}}_1$ $=$ $O [\Delta 
T]$, Eq. (\ref{hiroi}), and then the term $\lq\lq O\mbox{''} [g 
\Delta T]$ may be ignored. As mentioned above, the term $\tau 
{\bf p} \cdot \hat{{\bf q}}_1 \, (\underline{{\bf k}}_\perp)^2 / 
(2 q_1^2)$ in (\ref{yare-1}) is of $O [g^3]$ and, according to the 
preliminary remarks above after (\ref{hiroi}), seems to be ignored. 
It cannot be overemphasized, however, that {\em this term 
necessarily appears} in the combination {\em as in (\ref{yare-1})} 
and we keep it. 

From the analysis made so far, we may write $\displaystyle{ 
\raisebox{1.1ex}{\scriptsize{$(1)$}}} \mbox{\hspace{-0.3ex}} 
\hat{\Lambda}^{\mu = 0}_n$, the contribution from Fig. 6 with $n$ 
rungs, and $P_\mu \displaystyle{ 
\raisebox{1.1ex}{\scriptsize{$(1)$}}} \mbox{\hspace{-0.3ex}} 
\hat{\Lambda}^\mu_n$ as, 
\begin{eqnarray} 
& & \displaystyle{ \raisebox{1.1ex}{\scriptsize{$(1)$}}} 
\mbox{\hspace{-0.3ex}} \hat{\Lambda}^{\mu = 0}_n (Q_1, Q_2) 
\simeq {\cal G}_n (Q_1, Q_2) \, , \\ 
& & P_\mu \displaystyle{ \raisebox{1.1ex}{\scriptsize{$(1)$}}} 
\mbox{\hspace{-0.3ex}} \hat{\Lambda}^\mu_n (Q_1, Q_2) \nonumber \\ 
& & \mbox{\hspace*{3.8ex}} \simeq [E - \tau {\bf p} \cdot 
\hat{{\bf q}}_1 (1 - f_n) ] \, {\cal G}_n (Q_1, Q_2) + O \{ g^3 \} 
\, . 
\label{ino-2} 
\end{eqnarray} 
Here $f_n$ has come from $(\underline{{\bf k}}_\perp)^2 / (2 q_1^2)$ 
in (\ref{yare-1}), {\em which is positive for all $n$} and is of $O 
(g^2)$. Then, thanks to the first mean value theorem, we can safely 
assume that $f_n$ ($n = 1, 2, ...$) is of $O (g^2)$ and (at least 
$Re \, f_n$) is positive. Now we choose $f$ $\equiv$ $f_j$ such 
that, for all $i$ ($\neq j$), $Re \, f_j$ $\leq$ $Re \, f_i$. Eq. 
(\ref{yare-1}) with (\ref{sum-yo}) indicates that presumably $f_j$ 
$=$ $f_1$. Then, summing over $n$, we obtain 
\begin{eqnarray} 
& & \displaystyle{ \raisebox{1.1ex}{\scriptsize{$(1)$}}} 
\mbox{\hspace{-0.3ex}} \hat{\Lambda}^{\mu = 0} (Q_1, Q_2) 
\equiv \sum_{n = 1}^\infty \displaystyle{ 
\raisebox{1.1ex}{\scriptsize{$(1)$}}} \mbox{\hspace{-0.3ex}} 
\hat{\Lambda}^{\mu = 0}_n (Q_1, Q_2) \nonumber \\ 
& & \mbox{\hspace*{5ex}} \simeq {\cal G} (Q_1, Q_2) \, , 
\label{ino-11} \\ 
& & P_\mu \, \displaystyle{ \raisebox{1.1ex}{\scriptsize{$(1)$}}} 
\mbox{\hspace{-0.3ex}} \hat{\Lambda}^\mu (Q_1, Q_2) \equiv 
P_\mu \sum_{n = 1}^\infty \displaystyle{ 
\raisebox{1.1ex}{\scriptsize{$(1)$}}} \mbox{\hspace{-0.3ex}} 
\hat{\Lambda}^\mu_n (Q_1, Q_2) \nonumber \\ 
& & \mbox{\hspace*{5ex}} \simeq [E - \tau {\bf p} \cdot 
\hat{{\bf q}}_1 (1 - f) ] \, {\cal G} (Q_1, Q_2) + O \{ g^3 \} 
\, . 
\label{ino-22} 
\end{eqnarray} 
In obtaining (\ref{ino-22}), the term $\tau {\bf p} \cdot 
\hat{{\bf q}}_1 (f_n - f) \, {\cal G}_n$ in $P_\mu \displaystyle{ 
\raisebox{1.1ex}{\scriptsize{$(1)$}}} \mbox{\hspace{-0.3ex}} 
\hat{\Lambda}^\mu_n$ has been absorbed into the term $O \{ g^3 \}$ 
in (\ref{ino-2}) and then in (\ref{ino-22}). $f$ may depends on 
the (dropped) thermal indices. However, this dependence does not 
become an obstacle for our purpose (cf. next subsection). 

\begin{center} 
{\em The second term on the R.H.S. of (\ref{sindo-1})} 
\end{center} 

Let us turn to analyze the contribution, $\displaystyle{ 
\raisebox{1.1ex}{\scriptsize{$(2)$}}} \mbox{\hspace{-0.3ex}} 
\hat{\Lambda}^\mu (Q_1, Q_2)$, which arises from the second term on 
the R.H.S. of (\ref{sindo-1}): 
\begin{equation} 
{\cal E}^\mu \equiv \tau \, \gamma^\mu \, 
\hat{{Q\kern-0.1em\raise0.3ex\llap{/}\kern0.15em\relax}}_\tau \, 
\frac{1}{q} \left[ \vec{\gamma} \cdot \{{\bf p} - ({\bf p} \cdot 
\hat{{\bf q}}) \, \hat{{\bf q}}\} \right] \, . 
\label{mine-1} 
\end{equation} 
This is of $O (g)$ and does not contribute to $\hat{\Lambda}^{\mu 
= 0} (Q_1, Q_2)$ to leading order. Then it is sufficient to analyze 
the contribution to Ward-Takahashi relation. Multiplying $P_\mu (= 
Q_{2 \mu} - Q_{1 \mu})$ to ${\cal E}^\mu$ and summing over $\mu$, 
we obtain 
\begin{eqnarray*} 
& & \tau {P\kern-0.1em\raise0.3ex\llap{/}\kern0.15em\relax} \, 
\hat{{Q\kern-0.1em\raise0.3ex\llap{/}\kern0.15em\relax}}_{1 \tau} \, 
\frac{1}{q_1} \left[ \vec{\gamma} \cdot \{{\bf p} - ({\bf p} \cdot 
\hat{{\bf q}}_1) \, \hat{{\bf q}}_1 \} \right] \nonumber \\ 
& & \mbox{\hspace*{4ex}} \simeq \tau \left\{ \gamma^0 [E - \tau (q_2 
- q_1)] + \tau \, q_2 \, 
\hat{{Q\kern-0.1em\raise0.3ex\llap{/}\kern0.15em\relax}}_{2 \tau} 
\right\} \, 
\hat{{Q\kern-0.1em\raise0.3ex\llap{/}\kern0.15em\relax}}_{1 \tau} \, 
\frac{1}{q_1} \vec{\gamma} \cdot {\bf p}_\perp 
\nonumber \\ 
& & \mbox{\hspace*{6ex}} + O [g^{5 / 2}] \nonumber \\ 
& & \mbox{\hspace*{4ex}} \simeq \tau \left[ \gamma^0 (E - \tau 
{\bf p} \cdot \hat{{\bf q}}_1) - \vec{\gamma} \cdot {\bf p}_\perp 
\right] \, 
\hat{{Q\kern-0.1em\raise0.3ex\llap{/}\kern0.15em\relax}}_{1 \tau} \, 
\frac{1}{q_1} \vec{\gamma} \cdot {\bf p}_\perp \nonumber \\ 
& & \mbox{\hspace*{6ex}} + O [g^{5 / 2}] \nonumber \\ 
& & \mbox{\hspace*{4ex}} = O [g \Delta T] \times 
\hat{{Q\kern-0.1em\raise0.3ex\llap{/}\kern0.15em\relax}}_{1 \tau} 
+ O [g^{5 / 2}] \, . 
\end{eqnarray*} 
This is of the same order of magnitude as $O [g \Delta T]$ in 
(\ref{yare-1}) and we can ignore this contribution. 
\subsubsection{Nonladder diagram} 
Finally we make a brief analysis of nonladder diagram, Fig. 7, the 
contribution of which is of $O \{ g \}$ \cite{l-s}. Then, as in the 
case of (\ref{mine-1}), it is sufficient to analyze the contribution 
to Ward-Takahashi relation. As above, the dominant contribution 
comes from $|R^2|$, $|Q^2|$ $=$ $O [g^2 T^2]$, where $|{\bf p} - 
({\bf p} \cdot \hat{{\bf q}}) \, \hat{{\bf q}}|$ $=$ $O 
[(g \Delta)^{1 / 2} T]$ $\leq$ $O [g^{3 / 2} T]$. We pick out the 
term, 
\begin{equation} 
P_\mu \, 
\hat{{Q\kern-0.1em\raise0.3ex\llap{/}\kern0.15em\relax}}_\xi 
\gamma^\mu \, 
\hat{{R\kern-0.1em\raise0.3ex\llap{/}\kern0.15em\relax}}_\xi = 2 (P 
\cdot \hat{Q}_\xi) \, 
\hat{{R\kern-0.1em\raise0.3ex\llap{/}\kern0.15em\relax}}_\xi + ... 
\, , 
\label{non-1} 
\end{equation} 
where $\xi = \epsilon (q_0) = \epsilon (r_0)$. In (\ref{non-1}), 
\lq\lq $...$'' leads to the contribution (to $P_\mu 
\hat{\Lambda}^\mu (Q_1, Q_2)$), which is absorbed into the term $O 
\{ g^3 \}$ in (\ref{ino-22}) and may be ignored. The first term 
leads to 
\[ 
P_\mu \, \hat{\Lambda}^\mu (Q_1, Q_2) = \left( E - \tau {\bf p} 
\cdot \hat{{\bf q}}_1 + O [g^3] \right) \, {\cal G}' (Q_1, Q_2) 
\, . 
\] 
Then ${\cal G}' (= O \{ g \})$ here can be absorbed into ${\cal G}$ 
in (\ref{ino-22}).  
\subsubsection{The form for $\hat{\Lambda}^\mu (Q_1, Q_2)$} 
Through the qualitative analysis made above, we have learnt 
that the structures of $\hat{\Lambda}^{\mu = 0} (Q_1, Q_2)$ (to 
leading order) and $P_\mu \hat{\Lambda}^\mu (Q_1, Q_2)$ are given by 
(\ref{ino-11}) and (\ref{ino-22}), respectively: 
\begin{mathletters} 
\label{ino-ino} 
\begin{eqnarray} 
& & \hat{\Lambda}^{\mu = 0} (Q_1, Q_2) \simeq {\cal G} (Q_1, Q_2) 
\, , 
\eqnum{4.26a} 
\label{ino-111} \\ 
& & P_\mu \, \hat{\Lambda}^\mu (Q_1, Q_2) \nonumber \\ 
& & \mbox{\hspace*{3.6ex}} \simeq [E - \tau {\bf p} \cdot 
\hat{{\bf q}}_1 (1 - f_n) ] \, {\cal G} (Q_1, Q_2) + O \{ g^3 \} 
\, . 
\eqnum{4.26b} 
\label{ino-222} 
\end{eqnarray} 
\end{mathletters} 

We have to emphasize that the forms (\ref{ino-ino}) should not be 
taken too seriously. This is because (\ref{ino-ino}) is not the 
\lq\lq calculated'' result but is obtained by assuming the 
Ward-Takahashi relation supplemented with diagrammatic analysis. 
Nevertheless, to go further, we assume the forms (\ref{ino-111}) and 
(\ref{ino-222}) in the following. 

To (generalized) one-loop order, $\tilde{\Sigma} (Q)$ may be 
decomposed as 
\begin{equation} 
\tilde{\Sigma} (Q) \simeq \gamma^0 \, {\cal H}_0 (Q) + 
\hat{{Q\kern-0.1em\raise0.3ex\llap{/}\kern0.15em\relax}}_\tau \, 
{\cal H}_v (Q) \, , 
\label{de} 
\end{equation} 
where $q_0 \simeq \tau q$. Then, from (\ref{WT-ne}), 
(\ref{karage-2}) with (\ref{de}), we see that, ${\cal G} (Q_1, 
Q_2)$ in (\ref{ino-ino}) may be decomposed as 
\[ 
{\cal G} (Q_1, Q_2) \simeq \gamma^0 \, {\cal G}_0 (Q_1, Q_2) + 
\hat{{Q\kern-0.1em\raise0.3ex\llap{/}\kern0.15em\relax}}_{1 \tau} 
\, {\cal G}_v (Q_1, Q_2) \, . 
\] 
According to the preliminary remarks above after (\ref{hiroi}), the 
second term on the R.H.S. is not important for our purpose. 

Now we are ready to determine $[{\cal G}_0 (Q_1, Q_2)]^1_{12 / 21}$ 
or $[\hat{\Lambda}^{\mu = 0} (Q_1, Q_2)]^1_{12 / 21}$, to leading 
order. By substituting (\ref{ino-222}) into (\ref{WT-ne}) with 
(\ref{add}), we find 
\begin{eqnarray} 
\left( \hat{\Lambda}^{\mu = 0} (Q_1, Q_2) \right)^1_{12 / 21} 
& \simeq & \mp 2 i \tau \, \gamma^0 [\theta (\mp q_{1 0}) - n_F 
(q_1)] \nonumber \\ 
& & \times \frac{\gamma_q}{E - \tau {\bf p} \cdot \hat{{\bf q}}_1 
(1 - f)} \, . 
\label{4-30} 
\end{eqnarray} 
It should be emphasized that $Re \, f > 0$ and $|f| = O [g^2]$. Then 
(\ref{4-30}) is not singular at $|{\bf p} \cdot \hat{{\bf q}}_1|$. 
In (\ref{4-30}), we have set $\Gamma (q_1) = \gamma_q$, being valid 
at logarithmic accuracy and gauge independent, which is sufficient 
for our purpose. Also to be emphasized is the fact that the R.H.S. 
of (\ref{WT-ne}) with $j \neq i$ (and then also $\left( 
\hat{\Lambda}^{\mu = 0} \right)^1_{j \neq i}$) is independent of $Re 
\, \tilde{\Sigma}_F (Q)$ ($=$ $\epsilon (q_0) (m_f^2 / q) 
\gamma^0$). 

It is worth mentioning that, in determining the form of 
$\hat{\Lambda}^{\mu = 0}$, Eq. (\ref{4-30}), through the qualitative 
analysis, we have not used the explicit form of the soft-gluon 
propagator. Then, the \lq\lq qualitative'' result (\ref{4-30}) does 
not depend on the HTL-resummed (approximate) form for the soft-gluon 
propagator. 
\subsection{Contribution to the modified effective photon-quark 
vertex} 
We are now in a position to compute Fig. 8, where the photon-quark 
vertex with the square blob indicates $\hat{\Lambda}^\mu$, the 
zeroth component of which is given by (\ref{4-30}). The original 
effective photon-quark vertex, Fig. 2, is gauge independent and we 
are dealing with the modification of it near the light-cone, 
$(Q + K)^2$ $\simeq$ $(Q + K')^2$ $\simeq$ $0$. Then, as in Secs. II 
and III, we use the Feynman gauge for the hard-gluon propagator in 
Fig. 8: 
\begin{eqnarray} 
& & \left( \raisebox{1.1ex}{\scriptsize{$\diamond$}} 
\mbox{\hspace{0.1ex}} \hat{\Gamma}^\mu (K', K) \right)^1_{j i} 
\nonumber \\ 
& & \mbox{\hspace*{4ex}} = - i (-)^{i + j} g^2 \, C_F \int 
\frac{d^{\, 4} Q}{(2 \pi)^4} \, \gamma^\rho \, \displaystyle{ 
\raisebox{1.1ex}{\scriptsize{$\diamond$}}} \mbox{\hspace{-0.33ex}} 
S_{j \ell} (Q_1) \nonumber \\ 
& & \mbox{\hspace*{7ex}} \times \left[ \hat{\Lambda}^\mu (Q_1, Q_2) 
\right]^1_{\ell k} \, \displaystyle{ 
\raisebox{1.1ex}{\scriptsize{$\diamond$}}} 
\mbox{\hspace{-0.33ex}} S_{k i} (Q_2) \, \gamma_\rho \, \Delta_{i j} 
(Q) \, , \nonumber \\ 
\label{j-7} 
\end{eqnarray} 
where $Q_1 = Q + K'$ and $Q_2 = Q + K$. As in Secs. II and III and 
as will be seen below, we need only $\left( 
\raisebox{1.1ex}{\scriptsize{$\diamond$}} \mbox{\hspace{0.1ex}} 
\hat{\Gamma}^{\mu = 0} \right)^1_{j i}$ with $i \neq j$, to leading 
order at logarithmic accuracy. 

As in Sec. IIIB, for hard gluon line with $Q$ in Fig. 8, we are 
allowed to use the bare propagator, $\Delta_{i j} (Q)$ (cf. 
(\ref{j-7})). Straightforward manipulation using (\ref{4-30}) yields 
\begin{eqnarray} 
\left( \raisebox{1.1ex}{\scriptsize{$\diamond$}} 
\mbox{\hspace{0.1ex}} \hat{\Gamma}^{\mu = 0} (K', K) \right)^1_{1 2} 
& \simeq & - \frac{i g^2}{2} \, C_F \sum_{\tau = \pm} \tau \int 
\frac{d^{\, 3} q_1}{(2 \pi)^3} \, \frac{1}{q_1} \, 
\hat{{Q\kern-0.1em\raise0.3ex\llap{/}\kern0.15em\relax}}_{1 \tau} 
\nonumber \\ 
& & \times [\theta (\tau) + n_B (q_1)] [\theta (- \tau) - n_F (q_1)] 
\frac{\tilde{\gamma}_q}{1 - \tau \hat{{\bf p}} \cdot \hat{{\bf q}}_1 
(1 - f)} \nonumber \\ 
& & \times \sum_{\rho = \pm} \left[ \frac{\rho \{ \theta 
(\rho) - n_F (q_1) \}}{\hat{Q}_{1 \tau} \cdot K' - \tau m_f^2 / q_1 
+ i \rho \tau \gamma_q} \right. \nonumber \\ 
& & \left. \times \frac{1}{\hat{Q}_{1 \tau} \cdot K - \tau 
m_f^2 / q_1 - i \rho \tau \gamma_q} \right] \, , \nonumber \\ 
\label{j-8} 
\end{eqnarray} 
where $\tilde{\gamma}_q$ $(= O [g])$ is as in (\ref{til-gamma}). Let 
us pick out 
\begin{eqnarray} 
& & \int_{- 1}^1 d z \, \frac{\tilde{\gamma}_q}{1 - \tau 
\hat{{\bf p}} \cdot \hat{{\bf q}}_1 (1 - f)} \nonumber \\ 
& & \mbox{\hspace*{5ex}} \times \sum_{\rho = \pm} 
\left[ \frac{ \rho \{ \theta (\rho) - n_F (q_1) \}}{\left( 
\hat{Q}_{1 \tau} \cdot K' - \tau m_f^2 / q_1 + i \rho \tau \gamma_q 
\right) \left( \hat{Q}_{1 \tau} \cdot K - \tau m_f^2 / q_1 - i \rho 
\tau \gamma_q \right) } \right] \, , 
\label{j-9} 
\end{eqnarray} 
where $z \equiv - \tau \hat{{\bf p}} \cdot \hat{{\bf q}}$. We 
proceed as in Sec. III (cf. (\ref{ver-36}) - (\ref{ver-362})). 
Rewrite (\ref{j-9}) as 
\begin{eqnarray} 
\mbox{Eq. (\ref{j-9})} & = & \frac{\tilde{\gamma}_q}{E} \int_{- 1}^1 
\frac{d z}{1 + z (1 - f)} \sum_{\rho = \pm} \left[ \frac{ \rho \{ 
\theta (\rho) - n_F (q_1) \}}{1 + z - i \rho \tau \tilde{\gamma_q}} 
\right. \nonumber \\ 
& & \left. \times \left\{ \frac{1}{\hat{Q}_{1 \tau} \cdot K' - \tau 
m_f^2 / q_1 + i \rho \tau \gamma_q} - \frac{1}{\hat{Q}_{1 \tau} 
\cdot K - \tau m_f^2 / q_1 - i \rho \tau \gamma_q} \right\} \right] 
\, . \nonumber \\ 
\label{j-10} 
\end{eqnarray} 
\noindent The contribution of our interest comes from $z \simeq - 
1$ or $\hat{{\bf p}} \cdot \hat{{\bf q}}_1$ $\simeq \tau$. Then 
$\hat{Q}_{1 \tau} \cdot K'$ $\simeq$ $\hat{Q}_{1 \tau} \cdot K$. 
Thus we have 
\begin{eqnarray} 
\mbox{Eq. (\ref{j-10})} & \simeq & \frac{2 \pi}{E} \, 
\delta_{\gamma_q} (K \cdot \hat{P} - \tau m_f^2 / q_1) \nonumber \\ 
& & \times \left[ \{ 1 - n_F (q_1) \} \ln \left( \frac{- i \tau 
\tilde{\gamma}_q}{f} \right) \right. \nonumber \\ 
& & \left. + n_F (q_1) \ln \left( \frac{ i \tau 
\tilde{\gamma}_q}{f} \right) \right] \, . 
\label{j-11} 
\end{eqnarray} 
After all this, as in Sec. III, we may set $\delta_{\gamma_q} (K 
\cdot \hat{P} - \tau m_f^2 / q_1)$ $\to$ $\delta (K \cdot \hat{P})$ 
and, at logarithmic accuracy, we get 
\begin{eqnarray} 
\mbox{Eq. (\ref{j-11})} & \simeq & 2 \pi \, \ln \left( 
\frac{\tilde{\gamma}_q}{|f|} \right) \, \delta (K \cdot P) \nonumber 
\\ 
& \simeq & 2 \pi \, \ln (g^{- 1}) \, \delta (K \cdot P) \, . 
\label{j-12} 
\end{eqnarray} 

Substituting (\ref{j-12}) into (\ref{j-8}), we obtain 
\begin{eqnarray} 
\left( \raisebox{1.1ex}{\scriptsize{$\diamond$}} 
\mbox{\hspace{0.1ex}} \hat{\Gamma}^{\mu = 0} (K', K) \right)^1_{1 2} 
& \simeq & \frac{i \pi}{2} \, m_f^2 \, \ln ( g^{- 1}) \, 
\hat{{P\kern-0.1em\raise0.3ex\llap{/}\kern0.15em\relax}} \, 
\delta (K \cdot P) \nonumber \\ 
& = & \left( \raisebox{1.1ex}{\scriptsize{$\diamond$}} 
\mbox{\hspace{0.1ex}} \tilde{\Gamma}^{\mu = 0} (K', K) 
\right)^1_{1 2} \nonumber \\ 
& = & \left( \raisebox{1.1ex}{\scriptsize{$\diamond$}} 
\mbox{\hspace{0.1ex}} \tilde{\Gamma}^{\mu = 0} (K', K) 
\right)^1_{2 1} \, . 
\label{j-13} 
\end{eqnarray} 

Similar analysis yields 
\begin{equation} 
\left( \raisebox{1.1ex}{\scriptsize{$\diamond$}} 
\mbox{\hspace{0.1ex}} \hat{\Gamma}^{\mu = 0} (K', K) 
\right)^1_{2 1} \simeq \left( 
\raisebox{1.1ex}{\scriptsize{$\diamond$}} \mbox{\hspace{0.1ex}} 
\hat{\Gamma}^{\mu = 0} (K', K) \right)^1_{1 2} \, . 
\label{j-14} 
\end{equation} 
\subsection{Contribution to the soft-photon production rate} 
The contribution (to the soft-photon production rate) of our concern 
is (cf. (\ref{pil1})) 
\begin{eqnarray} 
E \, \frac{d W^{(\ell)}}{d^{\, 3} p} & = & \frac{e_q^2 e^2 N_c}{2 
(2 \pi)^3} \left[ g_{\mu 0} \hat{P}_\nu + g_{\nu 0} \hat{P}_\mu 
- \hat{P}_\mu \hat{P}_\nu \right] \nonumber \\ 
& & \times \int \frac{d^{\, 4} K}{(2 \pi)^4} \, tr \Big[ 
\displaystyle{ \raisebox{0.6ex}{\scriptsize{*}}} \! S_{i_1 i_4} (K) 
\left( \displaystyle{\raisebox{1.1ex}{\scriptsize{$\diamond$}}} 
\mbox{\hspace{0.1ex}} \Gamma^\nu (K', K) \right)^{2}_{i_4 i_3} \, 
\displaystyle{\raisebox{0.6ex}{\scriptsize{*}}} \! S_{i_3 i_2} (K') 
\left( \displaystyle{\raisebox{1.1ex}{\scriptsize{$\diamond$}}} 
\mbox{\hspace{0.1ex}} \Gamma^\mu (K', K) \right)^{1}_{i_2 i_1} \Big] 
\, , \nonumber \\ 
\label{k-1} 
\end{eqnarray} 
\noindent where 
\begin{equation} 
\left( \displaystyle{\raisebox{1.1ex}{\scriptsize{$\diamond$}}} 
\mbox{\hspace{0.1ex}} \Gamma^\mu \right)^\ell_{j k} = - (-)^\ell 
\delta_{\ell j} \, \delta_{\ell k} \, \gamma^\mu + \left( 
\displaystyle{\raisebox{1.1ex}{\scriptsize{$\diamond$}}} 
\mbox{\hspace{0.1ex}} \tilde{\Gamma}^\mu \right)^\ell_{j k} + 
\left( \displaystyle{\raisebox{1.1ex}{\scriptsize{$\diamond$}}} 
\mbox{\hspace{0.1ex}} \hat{\Gamma}^\mu \right)^\ell_{j k} \, . 
\label{k-2} 
\end{equation} 
$\displaystyle{\raisebox{1.1ex}{\scriptsize{$\diamond$}}} 
\mbox{\hspace{0.1ex}} \tilde{\Gamma}^\mu$ is the contribution from 
Fig. 4, which has been dealt with in Sec. IIIB (cf. (\ref{ver-362}) 
and (\ref{ver-38})), while $\displaystyle{ 
\raisebox{1.1ex}{\scriptsize{$\diamond$}}} \mbox{\hspace{0.1ex}} 
\hat{\Gamma}^\mu$ is the contribution from Fig. 8. $\displaystyle{ 
\raisebox{1.1ex}{\scriptsize{$\diamond$}}} \mbox{\hspace{0.1ex}} 
\tilde{\Gamma}^\mu + \displaystyle{ 
\raisebox{1.1ex}{\scriptsize{$\diamond$}}} \mbox{\hspace{0.1ex}} 
\hat{\Gamma}^\mu$ is given by (\ref{j-7}), where $\hat{\Lambda}^\mu 
(Q_1, Q_2)$ is replaced by $\Lambda^\mu (Q_1, Q_2)$, Eq. 
(\ref{add-t}). Then, (\ref{k-1}) includes the contribution 
(\ref{yuugenn}) obtained in Sec. III. 

Using (\ref{WT-ne}) with (\ref{add-t}), we find 
\begin{eqnarray} 
& & \displaystyle{ \raisebox{0.6ex}{\scriptsize{*}}} \! 
S_{j i_2} (K') \left( P_\mu \, \displaystyle{ 
\raisebox{1.1ex}{\scriptsize{$\diamond$}}} \mbox{\hspace{0.1ex}} 
\Gamma^\mu (K', K) \right)^\ell_{i_2 i_1} \, \displaystyle{ 
\raisebox{0.6ex}{\scriptsize{*}}} \! S_{i_1 i} (K) \nonumber \\ 
& & \mbox{\hspace*{4ex}} \simeq \delta_{\ell i} \, \displaystyle{ 
\raisebox{0.6ex}{\scriptsize{*}}} \! S_{j i} (K') - \delta_{\ell j} 
\, \displaystyle{ \raisebox{0.6ex}{\scriptsize{*}}} \! S_{j i} (K), 
\label{WT-n} 
\end{eqnarray} 
with no summation over $i$ and $j$. Here we have used the fact that, 
for our purpose, we can set $\displaystyle{ 
\raisebox{1.1ex}{\scriptsize{$\diamond$}}} \mbox{\hspace{-0.33ex}} 
S_{j i} (K)$ $\to$ $\displaystyle{ \raisebox{0.6ex}{\scriptsize{*}}} 
\! S_{j i} (K)$ (cf. observation made at the end of Sec. III). 

Using (\ref{WT-n}) and (\ref{k-2}), we obtain 
\begin{eqnarray} 
E \, \frac{d W^{(\ell)}}{d^{\, 3} p} & \simeq & 
\frac{2 e_q^2 e^2 N_c}{(2 \pi)^3} \, \frac{1}{E} \, Re \int 
\frac{d^{\, 4} K}{(2 \pi)^4} \, tr \left[ \displaystyle{ 
\raisebox{0.6ex}{\scriptsize{*}}} \! S_{2 i} (K') \left( 
\displaystyle{\raisebox{1.1ex}{\scriptsize{$\diamond$}}} 
\mbox{\hspace{0.1ex}} \tilde{\Gamma}^0  (K', K) \right)^{1}_{i 2} 
\right. \nonumber \\ 
& & \left. - \left( \displaystyle{ 
\raisebox{1.1ex}{\scriptsize{$\diamond$}}} \mbox{\hspace{0.1ex}} 
\tilde{\Gamma}^0  (K', K) \right)^{1}_{2 i} \, \displaystyle{ 
\raisebox{0.6ex}{\scriptsize{*}}} \! S_{i 2} (K) \right] \, , 
\label{k-3} 
\end{eqnarray} 
\noindent where use has been made of (\ref{j-13}) and (\ref{j-14}). 
Note that the first term on the R.H.S. of (\ref{k-2}), when 
substituted into (\ref{k-3}), does not yield the leading 
contribution {\em at logarithmic accuracy}. Comparison of 
(\ref{k-3}) with (\ref{pil22}), with $\displaystyle{ 
\raisebox{0.6ex}{\scriptsize{*}}} \Gamma^0 (K', K)$ $\to$ 
$\raisebox{1.1ex}{\scriptsize{$\diamond$}} \mbox{\hspace{0.1ex}} 
\tilde{\Gamma}^0 (K', K)$, shows that $E \, d W^{(\ell)} / d^{\, 3} 
p$ in (\ref{k-3}) is twice as large as (\ref{yuugenn}). 

Adding the hard contribution to (\ref{k-3}), we finally 
obtain 
\begin{eqnarray} 
E \, \frac{dW}{d^{\, 3} p} & \simeq & \frac{e_q^2 \, \alpha \, 
\alpha_s}{\pi^2} \, T^2 \left( \frac{m_f}{E} \right)^2 \, \ln 
(g^{- 1}) \ln \left( \frac{T}{m_f} \right) \nonumber \\ 
& \simeq & \frac{e_q^2 \, \alpha \, \alpha_s}{\pi^2} \, T^2 
\left( \frac{m_f}{E} \right)^2 \, \ln^2 (g^{- 1}) \, , 
\label{katai-2} 
\end{eqnarray} 
which is valid at logarithmic accuracy and is gauge independent. 
Half of $E d W / d^{\, 3} p$ above comes from the region $O [g]$ 
$\leq$ $1 - \hat{{\bf p}} \cdot \hat{{\bf q}}$ $<<$ $O 
(1)$ in $\displaystyle{ \raisebox{1.1ex}{\scriptsize{$\diamond$}}} 
\mbox{\hspace{0.1ex}} \tilde{\Gamma}^{\mu = 0} (K', K)$ in 
(\ref{3.20'}). The remaining half comes from the region $O [g^2]$ 
$\leq$ $1 - \tau \hat{{\bf p}} \cdot \hat{{\bf q}}_1$ $\leq$ $O [g]$ 
in $\displaystyle{ \raisebox{1.1ex}{\scriptsize{$\diamond$}}} 
\mbox{\hspace{0.1ex}} \hat{\Gamma}^{\mu = 0}  (K', K)$ in 
(\ref{j-8}) with (\ref{j-9}). This is the central result of this 
paper. 

There is one comment to make about the usage of $\displaystyle{ 
\raisebox{0.6ex}{\scriptsize{*}}} \! S$s in (\ref{k-1}). Just as in 
the photon-quark vertex dealt with in this section, corrections to 
the quark-gluon vertices in Fig. 5 should be taken into account. 
From the analysis in this section, it can readily be recognized that 
the corrections are important only in the region, $|k - |k_0||$ 
$\leq$ $O [g^2 T]$. Then, the same observation as above, made at the 
end of Sec. III, applies and the leading-order result 
(\ref{katai-2}) holds unchanged.   

Seemingly a correction to the photon-quark vertex as depicted in 
Fig. 9, where $Q_1$ and $Q_2$ are hard and all the three gluon lines 
carry soft momenta, is of $O (1)$, the same order of magnitude 
as the bare photon-quark vertex. In appendix C, we show that, as a 
matter of fact, the correction, Fig. 9, is at most of $O [g]$, so 
that it does not lead to leading contribution to the soft-photon 
production rate. It is straightforward to extend the analysis in 
Appendix C to more general diagrams for the photon-quark vertex, in 
which many soft-gluon lines participate. We then find that they do 
not yield the leading contribution to $\hat{\Lambda}^\mu 
(Q_1, Q_2)$. 
\setcounter{equation}{0}
\setcounter{section}{4}
\section{Discussions and conclusions} 
\def\theequation{\mbox{\arabic{section}.\arabic{equation}}}
In Sec. II, within the canonical HTL-resummation scheme, we have 
analyzed the diagram that lead to logarithmically divergent leading 
contribution to the soft-photon production rate. The diverging 
factor $1 / \hat{\epsilon}$ comes from mass singularity. As has been 
pointed out in \cite{reb,k-r-s,reb1}, if the calculation of some 
quantity within the HTL-resummation scheme results in a diverging 
result, it is a signal of necessity of further resummation. In Sec. 
III, by replacing the hard-quark propagators $S$s with 
$\displaystyle{ \raisebox{1.1ex}{\scriptsize{$\diamond$}}} 
\mbox{\hspace{-0.33ex}} S$s, which is obtained by resumming one-loop 
self-energy part $\tilde{\Sigma}_F$ in a self-consistent manner, we 
have shown that the mass singularity is screened and the diverging 
factor $1 / \hat{\epsilon}$ in the production rate turns out to $\ln 
(g^{- 1})$. Replacement $S$ $\to$ $\displaystyle{ 
\raisebox{1.1ex}{\scriptsize{$\diamond$}}} \mbox{\hspace{-0.33ex}} 
S$ violates the current-conservation condition, to which the 
photon-quark vertex subjects. In Sec. IV, we have estimated 
corrections to the photon-quark vertex, which inevitably comes in 
for restoring the current-conservation condition. The estimated 
corrections yield the leading contribution to the soft-photon 
production rate, which coincides with the contribution deduced in 
Sec. III. 
 
Thus, we have obtained for the soft-photon production rate, 
\begin{equation} 
E \, \frac{d W}{d^{\, 3} p} \simeq \frac{e_q^2 \, \alpha \, 
\alpha_s}{\pi^2} \, T^2 \left( \frac{m_f}{E} \right)^2 \, \ln^2 
(g^{- 1}) \, , 
\label{con-1} 
\end{equation} 
which is valid at logarithmic accuracy and is gauge independent. The 
result (\ref{con-1}) is twice as large as the result reported in 
\cite{reb1}, in which the \lq\lq asymptotic masses'' are resummed 
for hard propagators. In the present case, the asymptotic mass $m_f$ 
is in the denominators of $\displaystyle{ 
\raisebox{1.1ex}{\scriptsize{$\diamond$}}} \mbox{\hspace{-0.33ex}} 
\tilde{S}$s, Eqs. (\ref{saishuu-1}) - (\ref{sin-rho}), the mass 
which comes from the real part of the (hard) quark self-energy part, 
$Re \, \tilde{\Sigma}_F (R)$ $\simeq$ $\tau (m_f^2 / r) \gamma^0$, 
Eq. (\ref{yama-1}). However, at logarithmic accuracy, the term that 
should be kept in $\displaystyle{ 
\raisebox{1.1ex}{\scriptsize{$\diamond$}}} \mbox{\hspace{-0.33ex}} 
S$s is not $- \tau m_f^2 / r$ but the term including $\gamma_q$, the 
\lq\lq damping rate'', which comes from $Im \, \tilde{\Sigma}_F 
(R)$. 

Much interest has been devoted to the damping rate of moving quanta 
in a hot plasma \cite{pis1,damp,damp1,l-s}. Using 
(\ref{hajimari}), (\ref{tsugi}), and (\ref{con-1}), we obtain for 
the damping rate $\gamma$ of a soft photon in a quark-gluon plasma, 
\[ 
\gamma \simeq 2 \pi e_q^2 \, \alpha \, \alpha_s 
\, T \left( \frac{m_f}{E} \right)^2 \, \ln^2 (g^{- 1}) \, . 
\] 

Finally, we make general observation on the structure of a generic 
thermal reaction rate. Consider a generic formally higher-order 
diagram contributing to the reaction rate. Obviously, its 
contribution is really of higher order, in so far as that the 
loop-momentum integrations are carried out over the \lq\lq hard 
phase-space'' region. Here the \lq\lq hard phase-space region'' is 
the region where all the loop momenta $R$s are not only hard but 
also are \lq\lq hard'', $|R^\mu| >> O (g T)$ and $R^2 >> O 
( (g T)^2 )$. Then, the only possible source of emerging leading 
contributions from such diagrams are the infrared region and/or the 
region close to the light-cone in the loop-momentum space. 

As has already been mentioned at the end of Sec. IIIA, the mechanism 
of arising mass singularities due to hard propagators are the same 
as in vacuum theory \cite{kin,niemass}, since the statistical factor 
is finite for hard momentum. A mass singularity may arise from a 
collinear configuration of massless particles. Other parts of the 
diagram do not participate directly in the game. Then, as far as the 
mass-singular contributions (that turns out to \lq\lq $\ln 
(g^{- 1})$ contribution'') are concerned, we can use bare 
propagators for all but the hard lines that are responsible for mass 
singularities. In case of soft-photon production rate, dealt with in 
this paper, the hard gluon line in Figs. 2, 4, and 8 and hard lines 
constituting the HTL of the soft-quark self-energy part in Fig. 1 
are such propagators. After all, we see that using $\displaystyle{ 
\raisebox{1.1ex}{\scriptsize{$\diamond$}}} \mbox{\hspace{-0.33ex}} 
S$, $\displaystyle{ \raisebox{1.1ex}{\scriptsize{$\diamond$}}} 
\mbox{\hspace{-0.33ex}} \Delta^{\mu \nu}$, and $\displaystyle{ 
\raisebox{1.1ex}{\scriptsize{$\diamond$}}} \mbox{\hspace{-0.33ex}} 
\Delta^{(F P)}$, for, in respective order, the relevant hard quark, 
gluon, and FP-ghost propagators in the diverging (formally) 
higher-order diagram render the diverging contribution finite. 

As to the contribution from the infrared region or from the region 
where infrared region and the region close to the light-cone overlap 
each other, we cannot draw any definite conclusion at present. In 
fact, the structure of a generic thermal amplitude in such regions 
remains to be elucidated, the issue which is still under way 
\cite{reb,k-r-s,reb1}. We like to stress here that this issue is 
{\em not} inherent in the present case of the soft-photon 
production. In fact, any thermal reaction rate shares the same 
problem, even if it is finite to leading order. 
\section*{Acknowledgments} 
I sincerely thank R. Baier for useful discussions and criticisms at 
the early stages of this investigation. This work was supported in 
part by the Grant-in-Aide for Scientific Research ((A)(1) 
(No. 08304024)) of the Ministry of Education, Science and Culture of 
Japan. 
\setcounter{equation}{0}
\setcounter{section}{1}
\section*{Appendix A Thermal propagators}
\def\theequation{\mbox{\Alph{section}.\arabic{equation}}}
Here we display various expressions and useful formulas, which are 
directly used in this paper. 
\subsubsection*{Bare thermal propagator of a quark} 
\begin{eqnarray} 
S_{j \ell} (Q) & = & \sum_{\tau = \pm} 
\hat{{Q\kern-0.1em\raise0.3ex\llap{/}\kern0.15em\relax}}_\tau 
\tilde{S}^{(\tau)}_{j \ell} (Q) \, , 
\label{a-1} \\ 
\tilde{S}_{11}^{(\tau)} (Q) & = & - \left\{ \tilde{S}_{22}^{(\tau)} 
(Q) \right\}^* \, , \nonumber \\  
& = & \frac{1}{2 \{ q_0 (1 + i 0^+) - \tau q \}} \nonumber \\ 
& & + i \pi \epsilon (q_0) n_F (q) \, \delta (q_0 - \tau q) \, , 
\label{a-2} \\ 
\tilde{S}_{1 2 / 2 1}^{(\tau)} (Q) & = & \pm i \pi n_F (\pm q_0) \, 
\delta (q_0 - \tau q) \, . 
\label{a-3} 
\end{eqnarray}
\subsubsection*{Effective thermal propagator of a soft quark} 
\begin{equation}
\mbox{\hspace{-0.8ex}} 
\displaystyle{\raisebox{0.6ex}{\scriptsize{*}}} \! S_{j i} (K)
= \sum_{\sigma = \pm} 
\hat{{K\kern-0.1em\raise0.3ex\llap{/}\kern0.15em\relax}}_\sigma 
\displaystyle{\raisebox{0.6ex}{\scriptsize{*}}} \! 
\tilde{S}_{j i}^{(\sigma)} (K)\, , \mbox{\hspace{6ex}} 
(j, i = 1, 2) \, , 
\label{eff1}
\end{equation}

\noindent where 
\begin{eqnarray} 
\hat{K}_\sigma  & = & (1, \, \sigma \hat{{\bf k}}) \, , 
\nonumber \\ 
\displaystyle{\raisebox{0.6ex}{\scriptsize{*}}} \! 
\tilde{S}_{11}^{(\sigma)} (K) & = & - \left( 
\displaystyle{\raisebox{0.6ex}{\scriptsize{*}}} \! 
\tilde{S}_{22}^{(\sigma)} (K) \right)^* \nonumber \\ 
& = & - \, \frac{1}{2 D_\sigma (k_0 (1 + i 0^+), k)} \nonumber \\ 
& & +  i \pi \, \epsilon (k_0) n_F (|k_0|) \rho_\sigma (K) \, , 
\label{eff3} \\ 
\displaystyle{\raisebox{0.6ex}{\scriptsize{*}}} \! 
\tilde{S}_{1 2 / 2 1}^{(\sigma)} (K) & = & \pm i \pi \, 
n_F (\pm k_0) \, \mbox{$\large{\rho}_\sigma$} (K) \, , 
\label{eff4} 
\end{eqnarray} 

\noindent with
\begin{eqnarray}
D_\sigma (K) & = & - k_0 + \sigma k \nonumber \\ 
& & + \frac{m_f^2}{2 k} \left[ \left( 1 - \sigma 
\frac{k_0}{k} \right) \, \ln \frac{k_0 + k}{k_0 - k}  \, + \, 2 
\sigma \, \right] \, , 
\label{eff5} \\ 
\rho_\sigma (K) & = & \frac{\epsilon (k_0)}{2 \pi i} \left[ \, 
\frac{1}{D_\sigma (k_0 ( 1 + i 0^+) , k)} \right. \nonumber \\ 
& & \left. - \frac{1}{D_\sigma (k_0 ( 1- i 0^+), k)} \, \right] \, , 
\label{eff10} \\ 
m_f^2 & = & \frac{\pi \alpha_s}{2} \, \frac{N_c^2 - 1}{2 N_c} \, 
T^2 \, . \nonumber 
\end{eqnarray}

\noindent $D_\sigma (K)$ in (\ref{eff5}) is first calculated in 
\cite{wel}. 
\subsubsection*{Bare thermal gluon propagator} 
\begin{eqnarray} 
& & \Delta_{j \ell}^{\mu \nu} (Q) = - \left[ g^{\mu \nu} - \eta 
\, Q^\mu \, Q^\nu \frac{\partial}{\partial \lambda^2} \right] 
\, \Delta_{j \ell} (Q ; \lambda^2)
\rule[-3mm]{.14mm}{8.5mm} \raisebox{-2.85mm}{\scriptsize{$\; \lambda 
= 0$}} \, , \nonumber \\ 
& & \mbox{\hspace*{26ex}} (j, \, \ell = 1, 2) \, , 
\label{bosb1} \\ 
& & \Delta_{11} (Q ; \lambda^2) = - \left\{ \Delta_{22} 
(Q ; \lambda^2) \right\}^* \nonumber \\ 
& & \mbox{\hspace*{4ex}} = \frac{1}{Q^2 - \lambda^2 + i 0^+} - 2 \pi 
i n_B (|q_0|) \, \delta (Q^2 - \lambda^2) \, , 
\label{bosb3} \\ 
& & \Delta_{1 2 / 2 1} (Q ; \lambda^2) \nonumber \\ 
& & \mbox{\hspace*{4ex}} = - \, 2 \pi i \left[ \theta 
(\mp q_0) + n_B (|q_0|) \right] \, \delta (Q^2 - \lambda^2) \, . 
\label{bosb2} 
\end{eqnarray} 
\setcounter{equation}{0}
\setcounter{section}{2}
\section*{Appendix B Ladder diagram with had gluon exchange}
\def\theequation{\mbox{\Alph{section}.\arabic{equation}}}
In this Appendix, we shall show that Fig. 6 with hard $K_j$ 
($2 \leq j \leq n$) yields $O \{ g^2 \}$ contribution to 
$\hat{\Lambda}^\mu (Q_1, Q_2)$, while the contribution from hard 
$K_1$ is of the form $O (1) \times 
\hat{{Q\kern-0.1em\raise0.3ex\llap{/}\kern0.15em\relax}}_{1 \tau}$ 
$\simeq$ $O (1) \times 
\hat{{Q\kern-0.1em\raise0.3ex\llap{/}\kern0.15em\relax}}_{2 \tau}$. 
The analysis here is of \lq\lq minimum'' in the sense that we 
stop the analysis when we obtain the results that are sufficient for 
our purpose. 

Let $K_j$ in Fig. 6 be hard. From $\rule[-2mm]{.14mm}{6.5mm} \, 
\left( Q^{(j)}\right)^2 \rule[-2mm]{.14mm}{6.5mm} \, , \, 
\rule[-1.5mm]{.14mm}{5.5mm} \, \left( R^{(j)}\right)^2 
\rule[-1.5mm]{.14mm}{5.5mm} $ $<< T^2$ (cf. Sec. IVB), we can 
show that the important configurations are as follows: 
\begin{eqnarray} 
\begin{array}{lll} 
& \mbox{(a)} & \epsilon (k_j^0) = \epsilon_j: \nonumber \\ 
& & \mbox{\hspace*{4ex}} {\bf k}_j \;\, \mbox{and} \;\, 
{\bf q}^{(j)} \;\, \mbox{are nearly parallel,} \nonumber \\ 
& & \mbox{\hspace*{4ex}} \epsilon_{j + 1} = \epsilon_j \, , 
\nonumber \\ 
& & \mbox{\hspace*{4ex}} K_j^\mu / k_j \simeq Q^{(j) \, \mu} / 
q^{(j)} \nonumber \\ 
& & \mbox{\hspace*{10.8ex}} \simeq Q^{(j + 1) \, \mu} / q^{(j + 1)} 
\, . \nonumber \\ 
& \mbox{(b)} & \epsilon (k_j^0) = - \epsilon_j: \nonumber \\ 
& & \mbox{\hspace*{4ex}} {\bf k}_j \;\, \mbox{and} \;\, 
{\bf q}^{(j)} \;\, \mbox{are nearly antiparallel,} \nonumber \\ 
& & \mbox{\hspace*{4ex}} \epsilon_{j + 1} = \epsilon_j \, \epsilon 
(q^{(j)} - k_j) \, , \nonumber \\ 
& & \mbox{\hspace*{4ex}} K_j^\mu / k_j \simeq - Q^{(j) \, \mu} / 
q^{(j)} \nonumber \\ 
& & \mbox{\hspace*{10.8ex}} \simeq - 
\epsilon (q^{(j)} - k_j) \, Q^{(j + 1) \, \mu} / q^{(j + 1)} \, , 
\nonumber 
\end{array} 
\end{eqnarray} 
where $\epsilon_j \equiv \epsilon (q_0^{(j)})$. Taking this into 
account, we obtain 
\begin{eqnarray} 
q^{(j + 1)} & \simeq & |q^{(j)} + \zeta k_j| - \frac{\zeta q^{(j)} 
k_j (1 - \zeta \hat{{\bf q}}^{(j)} \cdot \hat{{\bf k}}_j)}{|q^{(j)} 
+ \zeta k_j|} \, , \\ 
\hat{Q}_{\epsilon_{j + 1}}^{(j + 1) \, \mu} & \simeq & 
\hat{Q}_{\epsilon_j}^{(j) \, \mu} + \epsilon_{j + 1} \left( 0, \, 
\underline{{\bf q}}^{(j)} \right)^\mu \, , 
\label{shitsu-0} \\ 
\underline{{\bf q}}^{(j)} & = & \frac{{\bf k}_j - \zeta k_j 
\hat{{\bf q}}^{(j)}}{|q^{(j)} + \zeta k_j|} \nonumber \\ 
& & + \frac{\epsilon_{j + 1} \, \epsilon (k_j^0) \, q^{(j)} \, 
k_j (1 - \zeta \hat{{\bf q}}^{(j)} \cdot \hat{{\bf k}}_j)}{|q^{(j)} 
+ \zeta k_j|^2} \, \hat{{\bf q}}^{(j)} \, , 
\label{shitsu-1} 
\end{eqnarray} 
where $\zeta \equiv \epsilon_j \, \epsilon (k_j^0)$ and 
$\hat{Q}_{\epsilon_j}^{(j) \, \mu} \simeq \hat{Q}_{1 \tau}^\mu$ or 
$- \hat{Q}_{1 \tau}^\mu$. 

The quark propagator $\displaystyle{ 
\raisebox{1.1ex}{\scriptsize{$\diamond$}}} \mbox{\hspace{-0.33ex}} 
\tilde{S}^{(\epsilon_{j + 1})} (Q^{(j + 1)})$, Eq. (\ref{pre-425}), 
may be written as 
\begin{equation} 
\displaystyle{ 
\raisebox{1.1ex}{\scriptsize{$\diamond$}}} \mbox{\hspace{-0.33ex}} 
\tilde{S}^{(\epsilon_{j + 1})} (Q^{(j + 1)}) \simeq \frac{1}{2} 
\sum_{\rho = \pm} \frac{{\cal N}^{(\rho)} 
(Q^{(j + 1)})}{{\cal D}_\rho} \, , 
\label{koi-2} 
\end{equation} 
where 
\begin{eqnarray} 
{\cal D}_\rho & = & q_0^{(j + 1)} - \epsilon_{j + 1} 
\overline{q}^{(j + 1)} + i \rho \epsilon_{j + 1} \, \gamma_q 
(q^{(j + 1)}) \nonumber \\ 
& \simeq & q_0^{(j)} + k^0_j - \epsilon_{j + 1} 
|q^{(j)} + \zeta k_j| - \epsilon_{j + 1} \, \frac{m_f^2}{|q^{(j)} + 
\zeta k_j|} \nonumber \\ 
& & + \epsilon (k_j^0) \, \frac{q^{(j)} \, k_j}{q^{(j)} + \zeta k_j} 
\, (1 - \zeta \hat{{\bf q}}^{(j)} \cdot \hat{{\bf k}}_j) \nonumber 
\\ 
& & + i \rho \, \epsilon_{j + 1} \gamma_q (q^{(j + 1)}) \, . 
\label{koi-3} 
\end{eqnarray} 
As has been mentioned in Sec. IVB, the leading contribution to 
$\hat{\Lambda}^\mu (Q_1, Q_2)$ comes from the region; 
$\rule[-2mm]{.14mm}{6.5mm} \, \left( Q^{(j)} \right)^2 
\rule[-2mm]{.14mm}{6.5mm} \,$, $\rule[-2mm]{.14mm}{6.5mm} \, 
\left( Q^{(j + 1)} \right)^2 \rule[-2mm]{.14mm}{6.5mm} \,$, 
$|K_j^2|$ $\leq$ $O [g^2 T^2]$. Then from (\ref{koi-2}) and 
(\ref{koi-3}), we see that the important region is 
\begin{eqnarray} 
& & 1 - \zeta \, \hat{{\bf q}}^{(j)} \cdot \hat{{\bf k}}_j \, , 
\;\;\; 1 - \zeta \, \hat{{\bf r}}^{(j)} \cdot \hat{{\bf k}}_j 
\leq  O [g^2] \, , \nonumber \\ 
& & |k_j^0 - \epsilon (k_j^0) \, k_j| \leq O [g^2 T] \, . 
\label{b-1} 
\end{eqnarray} 
The region, e.g., $1 - \zeta \, \hat{{\bf q}}^{(j)} \cdot 
\hat{{\bf k}}_j$ $=$ $O [g]$ yields $O \{ g \}$ contribution to 
$\hat{\Lambda}^\mu (Q_1, Q_2)$. 

The hard-gluon propagator $\displaystyle{ 
\raisebox{1.1ex}{\scriptsize{$\diamond$}}} \mbox{\hspace{-0.33ex}} 
\Delta^{\rho \sigma} (K)$ consists \cite{sin-nie} of three part, the 
transverse part, the longitudinal part, and the gauge part. Their 
Lorentz-tensor structure are ${\cal P}_T^{\rho \sigma} 
(\hat{{\bf k}})$ $\equiv$ $- \sum_{i, \, j = 1}^3 g^{\rho i} 
g^{\sigma j} (\delta^{i j} - \hat{k}^{i} \hat{k}^{j})$, ${\cal 
P}_L^{\rho \sigma} (K)$ $\equiv$ $g^{\rho \sigma}$ $-$ $K^{\rho} 
K^{\sigma} / K^{2}$ $-$ ${\cal P}_T^{\rho \sigma} (\hat{{\bf k}})$, 
and $K^{\rho} K^{\sigma} / K^2$, respectively. 

Now, from Fig. 6, we pick out 
\begin{eqnarray} 
{\cal I}_Q^{\sigma_j} & \equiv & \gamma^{\sigma_j} \,  
\hat{{Q\kern-0.1em\raise0.3ex\llap{/}\kern0.15em\relax}}_{
\epsilon_{j + 1}}^{(j + 1)} \nonumber \\ 
& = & - \hat{{Q\kern-0.1em\raise0.3ex\llap{/}\kern0.15em\relax}}_{
\epsilon_{j + 1}}^{(j + 1)} \, \gamma^{\sigma_j} + 2 
\hat{Q}_{\epsilon_{j + 1}}^{(j + 1) \, \sigma_j} \nonumber \\ 
& \equiv & {\cal I}_Q^{(1) \, \sigma_j} + {\cal I}_Q^{(2) \, 
\sigma_j} \, , \nonumber \\ 
{\cal I}_R^{\rho_j} & \equiv & 
\hat{{R\kern-0.1em\raise0.3ex\llap{/}\kern0.15em\relax}}_{
\epsilon_{j + 1}}^{(j + 1)} \, \gamma^{\rho_j} \, , \nonumber \\ 
& = & - \gamma^{\rho_j} \, 
\hat{{R\kern-0.1em\raise0.3ex\llap{/}\kern0.15em\relax}}_{
\epsilon_{j + 1}}^{(j + 1)} + 2 \hat{R}_{\epsilon_{j + 1}}^{(j + 1) 
\, \rho_j} \nonumber \\ 
& \equiv & {\cal I}_R^{(1) \, \rho_j} + {\cal I}_R^{(2) \, \rho_j} 
\, . 
\label{uso-1} 
\end{eqnarray}
\subsubsection{${\cal I}_Q^{(1)} \otimes {\cal I}_R^{(1)}$} 
We start with analyzing ${\cal I}^{(1) \, \sigma_j}_Q \otimes 
{\cal I}^{(1) \, \rho_j}_R$. Using (\ref{shitsu-0}) and 
(\ref{shitsu-1}), we obtain 
\begin{eqnarray} 
{\cal I}_Q^{(1) \, \sigma_j} & = & - \left[ 
\hat{{Q\kern-0.1em\raise0.3ex\llap{/}\kern0.15em\relax}}_{
\epsilon_j}^{(j)} - \epsilon_{j + 1} \vec{\gamma} \cdot 
\underline{{\bf q}}^{(j)} \right] \, \gamma^{\sigma_j} \, , 
\label{b-3} \\ 
{\cal I}_R^{(1) \, \rho_j} & = & - \gamma^{\rho_j} \, \left[ 
\hat{{R\kern-0.1em\raise0.3ex\llap{/}\kern0.15em\relax}}_{
\epsilon_j}^{(j)} - \epsilon_{j + 1} \vec{\gamma} \cdot 
\underline{{\bf r}}^{(j)} \right] \, , 
\label{b-4} 
\end{eqnarray} 
where $\underline{{\bf q}}^{(j)}$ is as in (\ref{shitsu-1}) and is 
of $O [g]$ in the region (\ref{b-1}). $\underline{{\bf r}}^{(j)}$ 
is defined by (\ref{shitsu-1}) with $Q^{(j)}$ $\to$ $R^{(j)}$. 

For $2 \leq j \leq n$, 
$\hat{{Q\kern-0.1em\raise0.3ex\llap{/}\kern0.15em\relax}}_{
\epsilon_j}^{(j)}$ 
$\left[ \hat{{R\kern-0.1em\raise0.3ex\llap{/}\kern0.15em\relax}}_{
\epsilon_j}^{(j)} \right]$ is to be multiplied from the left [right] 
of (\ref{b-3}) [(\ref{b-4})] and then the first terms in the square 
brackets in (\ref{b-3}) and (\ref{b-4}) vanish. Then, 
\[ 
\hat{{Q\kern-0.1em\raise0.3ex\llap{/}\kern0.15em\relax}}_{
\epsilon_j}^{(j)} \, {\cal I}_Q^{(1) \, \sigma_j} \otimes 
{\cal I}_R^{(1) \, \rho_j} \, 
\hat{{R\kern-0.1em\raise0.3ex\llap{/}\kern0.15em\relax}}_{
\epsilon_j}^{(j)} \leq O [g^2] \, . 
\] 
According to the preliminary remarks in Sec. IVB, this contribution 
can be ignored. 

For $j = 1$, we have 
\begin{eqnarray} 
{\cal I}_Q^{(1) \, \sigma_1} \otimes {\cal I}_R^{(1) \, \rho_1} 
& = & \hat{{Q\kern-0.1em\raise0.3ex\llap{/}\kern0.15em\relax}}_{1 
\tau} \, \gamma^{\sigma_1} \, \otimes \gamma^{\rho_1} \, \, 
\hat{{Q\kern-0.1em\raise0.3ex\llap{/}\kern0.15em\relax}}_{2 \tau} 
\nonumber \\ 
& & - \epsilon_2 \{ \vec{\gamma} \cdot \underline{{\bf q}}^{(1)} \} 
\gamma^{\sigma_1} \otimes \gamma^{\rho_1} \, 
\hat{{Q\kern-0.1em\raise0.3ex\llap{/}\kern0.15em\relax}}_{2 \tau} 
\nonumber \\ 
& & - \epsilon_2 
\hat{{Q\kern-0.1em\raise0.3ex\llap{/}\kern0.15em\relax}}_{1 \tau} 
\, \gamma^{\sigma_1} \otimes \gamma^{\rho_1} \, \{ \vec{\gamma} 
\cdot \underline{{\bf r}}^{(1)} \} + O [g^2] \, . \nonumber \\ 
\label{b-10} 
\end{eqnarray} 
To be consistent with the Ward-Takahashi relation (\ref{WT-ne}), the 
$4 \times 4$ matrix structure of $P_\mu \hat{\Lambda}^\mu 
(Q_1, Q_2)$ should be of the form (cf. (\ref{de})), 
\begin{eqnarray} 
P_\mu \hat{\Lambda}^\mu (Q_1, Q_2) & = & \gamma^0 \, {\cal F}_0 
(Q_1, Q_2) + 
\hat{{Q\kern-0.1em\raise0.3ex\llap{/}\kern0.15em\relax}}_{1 \tau} \, 
{\cal F}_1 (Q_1, Q_2) \nonumber \\ 
& & + \hat{{Q\kern-0.1em\raise0.3ex\llap{/}\kern0.15em\relax}}_{2 
\tau} \, {\cal F}_2 (Q_1, Q_2) \, . 
\label{b-11} 
\end{eqnarray} 
Comparison of (\ref{b-10}) and (\ref{b-11}) tells us that the 
contribution from ${\cal I}_Q^{(1) \, \sigma_1} \otimes 
{\cal I}_R^{(1) \, \rho_1}$ to $\hat{\Lambda}^\mu (Q_1, Q_2)$ is of 
the form 
\[ 
\hat{\Lambda}^\mu (Q_1, Q_2) = O (1) \times 
\hat{{Q\kern-0.1em\raise0.3ex\llap{/}\kern0.15em\relax}}_{1 \tau} + 
O (1) \times 
\hat{{Q\kern-0.1em\raise0.3ex\llap{/}\kern0.15em\relax}}_{2 \tau} + 
O [ g^2 ] \, . 
\] 
Here the first two contributions on the R.H.S. come from Fig. 6, 
where all $K$s but $K_1$ are soft. 
\subsubsection{${\cal I}_Q^{(1)} \otimes {\cal I}_R^{(2)}$ and 
${\cal I}_Q^{(2)} \otimes {\cal I}_R^{(1)}$} 
Let us turn to analyze ${\cal I}_Q^{(1) \, \sigma_j} \otimes 
{\cal I}_R^{(2) \, \rho_j}$. The first entry to be analyzed is 
\begin{eqnarray} 
{\cal I}_Q^{(1) \, \sigma_j} \otimes {\cal I}_R^{(2) \, \rho_j} \, 
{\cal P}_{T \, \rho_j \sigma_j} (\hat{{\bf k}}_j) & \ni & \left[ 
\hat{{Q\kern-0.1em\raise0.3ex\llap{/}\kern0.15em\relax}}_{
\epsilon_j}^{(j)} - \epsilon_{j + 1} \vec{\gamma} \cdot 
\underline{{\bf q}}^{(j)} \right] \, \gamma^{\sigma_j} \, 
\hat{R}_{\epsilon_{j + 1}}^{(j + 1) \, \rho_j} \, {\cal P}_{T \, 
\rho_j \sigma_j} (\hat{{\bf k}}_j) \nonumber \\ 
& = & - \epsilon_{j + 1} \left[ 
\hat{{Q\kern-0.1em\raise0.3ex\llap{/}\kern0.15em\relax}}_{
\epsilon_j}^{(j)} - \epsilon_{j + 1} \vec{\gamma} \cdot 
\underline{{\bf q}}^{(j)} \right] \nonumber \\ 
&& \times \left[ \vec{\gamma} \cdot 
\hat{{\bf r}}^{(j + 1)} - (\vec{\gamma} \cdot \hat{{\bf k}}_j) 
(\hat{{\bf k}}_j \cdot \hat{{\bf r}}^{(j + 1)} ) \right] \, . 
\label{b-19} 
\end{eqnarray} 
\noindent We see from (\ref{b-1}) with $j \to j + 1$ that the 
quantity in the second curly brackets in (\ref{b-19}) is of $O [g]$. 
Then, for $2 \leq j \leq n$, we have 
\[ 
\hat{{Q\kern-0.1em\raise0.3ex\llap{/}\kern0.15em\relax}}_{
\epsilon_j}^{(j)} \times \mbox{(\ref{b-19})} \leq O [g^2] 
\] 
and, for $j = 1$, 
\[ 
\mbox{(\ref{b-19})} \leq 
\hat{{Q\kern-0.1em\raise0.3ex\llap{/}\kern0.15em\relax}}_{1 \tau} 
\times O [g] + O [g^2] \, . 
\] 

The second entry is 
\begin{eqnarray} 
& & {\cal I}_Q^{(1) \, \sigma_j} \otimes {\cal I}_R^{(2) \, \rho_j} 
\, g_{\rho_j \sigma_j} \nonumber \\ 
& & \mbox{\hspace*{4ex}} \ni \left[ 
\hat{{Q\kern-0.1em\raise0.3ex\llap{/}\kern0.15em\relax}}_{
\epsilon_j}^{(j)} - \epsilon_{j + 1} \vec{\gamma} \cdot 
\underline{{\bf q}}^{(j)} \right] \, \gamma^{\sigma_j} \, 
\hat{R}_{\epsilon_{j + 1}}^{(j + 1) \, \rho_j} \, g_{\rho_j 
\sigma_j} \nonumber \\ 
& & \mbox{\hspace*{4ex}} = \left[ 
\hat{{Q\kern-0.1em\raise0.3ex\llap{/}\kern0.15em\relax}}_{
\epsilon_j}^{(j)} - \epsilon_{j + 1} \vec{\gamma} \cdot 
\underline{{\bf q}}^{(j)} \right] \, 
\hat{{R\kern-0.1em\raise0.3ex\llap{/}\kern0.15em\relax}}_{
\epsilon_{j + 1}}^{(j + 1)} \, . 
\label{b-40} 
\end{eqnarray} 
From ${\bf r}^{(j + 1)} = {\bf q}^{(j + 1)} + {\bf p}$, we obtain 
\begin{equation} 
\hat{{\bf r}}^{(j + 1)} \simeq \hat{{\bf q}}^{(j + 1)} + 
\frac{1}{q^{(j + 1)}} \left[ {\bf p} - ({\bf p} \cdot 
\hat{{\bf q}}^{(j + 1)}) \, \hat{{\bf q}}^{(j + 1)} \right] \, . 
\label{b-20} 
\end{equation} 
In the region (\ref{hiroi}) and (\ref{b-1}), the second term on the 
R.H.S. is of $O [(g \Delta)^{1 / 2}]$ $\leq$ $O [g^{3 / 2}]$. Then, 
using (\ref{shitsu-0}) with (\ref{shitsu-1}) and 
$(\hat{Q}_{\epsilon_j}^{(j)})^2 = 0$, we see that, in (\ref{b-40}), 
$\hat{{Q\kern-0.1em\raise0.3ex\llap{/}\kern0.15em\relax}}_{
\epsilon_j}^{(j)} \, 
\hat{{R\kern-0.1em\raise0.3ex\llap{/}\kern0.15em\relax}}_{
\epsilon_{j + 1}}^{(j + 1)}$ $=$ 
$\hat{{Q\kern-0.1em\raise0.3ex\llap{/}\kern0.15em\relax}}_{
\epsilon_j}^{(j)} \times O [g]$. The remainder of (\ref{b-40}) 
becomes 
\[ 
2 \underline{{\bf q}}^{(j)} \cdot \hat{{\bf r}}^{(j + 1)} + 
\epsilon_{j + 1} \, 
\hat{{R\kern-0.1em\raise0.3ex\llap{/}\kern0.15em\relax}}_{
\epsilon_{j + 1}}^{(j + 1)} \, \vec{\gamma} \cdot 
\underline{{\bf q}}^{(j)} \, . 
\] 
Eqs. (\ref{shitsu-1}) and (\ref{b-20}) tell us that the first term 
on the R.H.S is of $O [g^2]$ in the region (\ref{b-1}). The second 
term is of the form 
\[ 
\hat{{R\kern-0.1em\raise0.3ex\llap{/}\kern0.15em\relax}}_{
\epsilon_j}^{(j)} \times O [g] + O [g^2] = 
\hat{{Q\kern-0.1em\raise0.3ex\llap{/}\kern0.15em\relax}}_{
\epsilon_j}^{(j)} \times O [g] + O [g^2] \, . 
\] 
Thus, we have 
\[ 
\mbox{Eq. (\ref{b-40})} = 
\hat{{Q\kern-0.1em\raise0.3ex\llap{/}\kern0.15em\relax}}_{
\epsilon_j}^{(j)} \times O [g] + O [g^2] \, . 
\] 
For $2 \leq j \leq n$, 
$\hat{{Q\kern-0.1em\raise0.3ex\llap{/}\kern0.15em\relax}}_{
\epsilon_j}^{(j)} \times \mbox{(\ref{b-40})}$ $=$ $O [g^2]$. 

The third entry is 
\begin{eqnarray} 
& & {\cal I}_Q^{(1) \, \sigma_j} \otimes {\cal I}_R^{(2) \, \rho_j} 
\, K_{j \rho_j} \, K_{j \sigma_j} \nonumber \\ 
& & \mbox{\hspace*{4ex}} \ni \left[ 
\hat{{Q\kern-0.1em\raise0.3ex\llap{/}\kern0.15em\relax}}_{
\epsilon_j}^{(j)} - \epsilon_{j + 1} \vec{\gamma} \cdot 
\underline{{\bf q}}^{(j)} \right] \, \gamma^{\sigma_j} \, 
\hat{R}_{\epsilon_{j + 1}}^{(j + 1) \, \rho_j} \, K_{j \rho_j} \, 
K_{j \sigma_j} \nonumber \\ 
& & \mbox{\hspace*{4ex}} = \left[ 
\hat{{Q\kern-0.1em\raise0.3ex\llap{/}\kern0.15em\relax}}_{
\epsilon_j}^{(j)} - \epsilon_{j + 1} \vec{\gamma} \cdot 
\underline{{\bf q}}^{(j)} \right] \, 
\hat{{K\kern-0.1em\raise0.3ex\llap{/}\kern0.15em\relax}}_j \, 
(\hat{R}_{\epsilon_{j + 1}}^{(j + 1)} \cdot K_j) \, . 
\label{b-50} 
\end{eqnarray} 
In the region (\ref{b-1}), $K_j = \epsilon (k_j^0) \, k_j \, 
\hat{K}_j + O [g^2 T]$, where $\hat{K}_j = (1, \epsilon (k_j^0) \, 
\hat{{\bf k}}_j)$. Then, we obtain 
\begin{eqnarray} 
\hat{R}^{(j + 1)}_{\epsilon_{j + 1}} \cdot K_j & = & \epsilon 
(k_j^0) \, k_j \, \hat{R}_{\epsilon_{j + 1}}^{(j + 1)} \cdot 
\hat{K}_j + O [g^2 T] \nonumber \\ 
& = & \epsilon (k_j^0) \, k_j \left[ 1 - \epsilon_{j + 1} \, 
\epsilon (k_j^0) \, \hat{{\bf k}}_j \cdot \hat{{\bf r}}^{(j 
+ 1)} \right] \nonumber \\ 
& & + O [g^2 T] \nonumber \\ 
& = & O [g^2 T ] \, , 
\label{b-100} 
\end{eqnarray} 
where use has been made of (\ref{b-20}), (\ref{shitsu-0}), 
(\ref{shitsu-1}), and (\ref{b-1}). The remainder of (\ref{b-50}) may 
be analyzed as in the second entry above and we obtain 
\[ 
\mbox{Eq. (\ref{b-50})} = 
\hat{{Q\kern-0.1em\raise0.3ex\llap{/}\kern0.15em\relax}}_{
\epsilon_j}^{(j)} \times O [g^3 T^2] + O [g^4 T^2]  \, . 
\] 
We recall that the denominator of the hard-gluon propagator that 
accompanies to (\ref{b-50}) is $O [g^2 T^2]$ smaller than those 
accompanying to (\ref{b-10}) and (\ref{b-40}) above (cf. 
(\ref{bosb1}) - (\ref{bosb2})). 

${\cal I}_Q^{(2) \, \sigma_j} \otimes {\cal I}_R^{(1) \, \rho_j}$, 
Eq. (\ref{uso-1}), may be analyzed similarly as in the case of 
${\cal I}_Q^{(1) \, \sigma_j} \otimes {\cal I}_R^{(2) \, \rho_j}$ 
and the same conclusion results. 
\subsubsection{${\cal I}_Q^{(2)} \otimes {\cal I}_R^{(2)}$} 
Finally we analyze ${\cal I}_Q^{(2) \, \sigma_j} \otimes 
{\cal I}_R^{(2) \, \rho_j}$ in (\ref{uso-1}). The first entry is 
\begin{eqnarray*} 
& & {\cal I}_Q^{(2) \, \sigma_j} \, {\cal I}_R^{(2) \, \rho_j} \, 
{\cal P}_{T \, \rho_j \sigma_j} (\hat{{\bf k}}_j) \\ 
& & \mbox{\hspace*{4ex}} \ni \hat{{\bf q}}^{(j + 1)} \cdot 
\hat{{\bf r}}^{(j + 1)} - (\hat{{\bf q}}^{(j + 1)} \cdot \hat{{\bf 
k}}_j) (\hat{{\bf k}}_j \cdot \hat{{\bf r}}^{(j + 1)}) \nonumber \\ 
& & \mbox{\hspace*{4ex}} = O [g^2] \, . 
\end{eqnarray*} 

The second entry is 
\begin{eqnarray*} 
{\cal I}_Q^{(2) \, \sigma_j} \, {\cal I}_R^{(2) \, \rho_j} \, 
g_{\rho_j \sigma_j} & \ni & \hat{Q}^{(j + 1)}_{\epsilon_{j + 1}} 
\cdot \hat{R}^{(j + 1)}_{\epsilon_{j + 1}} \nonumber \\ 
& = & 1 - \hat{{\bf q}}^{(j + 1)} \cdot \hat{{\bf r}}^{(j + 1)} \, , 
\end{eqnarray*} 
which, according to (\ref{b-20}), is negligibly small. 

The third entry is 
\begin{eqnarray*} 
{\cal I}_Q^{(2) \, \sigma_j} \, {\cal I}_R^{(2) \, \rho_j} \, 
K_{j \rho_j} \, K_{j \sigma_j} & \ni & (\hat{Q}^{(j + 
1)}_{\epsilon_{j + 1}} \cdot K_j) (\hat{R}^{(j + 1)}_{\epsilon_{j + 
1}} \cdot K_j) \nonumber \\ 
& = & O [g^4 T^2] \, , 
\end{eqnarray*} 
where use has been made of (\ref{b-100}). 

This completes the proof of the statement made at the beginning of 
this Appendix. 
\setcounter{equation}{0}
\setcounter{section}{3}
\section*{Appendix C Brief analysis of Fig. 9}
\def\theequation{\mbox{\Alph{section}.\arabic{equation}}}
Here we briefly analyze Fig. 9, where $Q_1$ $(= Q + K')$ and $Q_2$ 
$(= Q + K)$ are hard, all the three gluon lines carry soft momenta, 
$K_1$, $K_2$, and $K_3$ $(= - K_1 - K_2)$, and show that its 
contribution is nonleading. 

The effective soft-gluon propagator, $\displaystyle{ 
\raisebox{0.6ex}{\scriptsize{*}}} \! \Delta_{i j}^{\xi \zeta} (K)$, 
consists of two terms, the one is proportional to $n_B (k_0)$ 
$\simeq$ $T / k_0$ ($= O (1 / g)$)and the other is independent of 
$n_B (k_0)$. The former term is of $O (1 / (g^3 T^2) )$, while the 
latter term is of $O (1 / (g^2 T^2))$. As in the case of bare 
thermal propagator, the former term is independent of thermal 
indices, $i$ and $j$. 

For a given set of thermal indices $(i_1 - i_3, j_1 - j_3)$ in Fig. 
9, we assign, on trial, the leading part of $\displaystyle{ 
\raisebox{0.6ex}{\scriptsize{*}}} \! \Delta_{i j}^{\xi \zeta}$ 
($\sim O (1 / (g^3 T^2))$) to the three gluon propagators. It can be 
shown explicitly that no divergence arises. In similar manner as in 
Sec. IV and Appendix B, we can estimate the order of magnitude of 
other characters in Fig. 9: $\displaystyle{ 
\raisebox{1.1ex}{\scriptsize{$\diamond$}}} \mbox{\hspace{-0.33ex}} 
S^{(\tau)}_{1 i_1} (R + K_1)$, $\displaystyle{ 
\raisebox{1.1ex}{\scriptsize{$\diamond$}}} \mbox{\hspace{-0.33ex}} 
S^{(\tau)}_{i_3 1} (Q + K_1)$, and $\displaystyle{ 
\raisebox{1.1ex}{\scriptsize{$\diamond$}}} \mbox{\hspace{-0.33ex}} 
S^{(\tau)}_{i_1 i_3} (Q - K_2)$ are of $O (1 / \tilde{\gamma}_q)$. 
A tri-gluon vertex is of $O (g T)$. $\int d^{\, 4} K_1$ $=$ $\int 
d k_{1 0} \, d k_1 \, k_1^2 \, d (\hat{{\bf k}}_1 \cdot 
\hat{{\bf q}})$ $=$ $O \{(g T)^4 \} \times O \{ \tilde{\Gamma} / 
(g T) \}$ $=$ $O \{ (g T)^3 \tilde{\Gamma}\}$ $=$ $\int d^{\, 4} 
K_2$. After all this, we see that, aside from a possible factor of 
$\ln (g^{- 1})$, Fig. 9 is of $O (1)$, the same order of magnitude 
as the bare photon-quark vertex. 

Now we note that, as mentioned above, the leading part of 
$\displaystyle{\raisebox{0.6ex}{\scriptsize{*}}} \! \Delta_{i 
j}^{\xi \zeta} (K)$ is independent of the thermal indices and the 
three-gluon vertex with a blob in Fig. 9 may be written as 
\[ 
{\cal V}_{j_1 j_2 j_3} = g \left[ (-)^{j_1} \delta_{j_1 j_2} 
\delta_{j_1 j_3} {\cal V}^{(0)} + 
{\cal V}^{\mbox{\scriptsize{(HTL)}}}_{j_1 j_2 j_3} \right] \, , 
\] 
where the Lorentz indices are deleted. The first term comes from the 
bare vertex and the second term represents the HTL contribution. 
Recalling the identity $\sum_{j_1, j_2, j_3 = 1}^2 
{\cal V}^{\mbox{\scriptsize{(HTL)}}}_{j_1 j_2 j_3} = 0$, we see 
that, upon summation over thermal indices, $j_1, j_2$ and $j_3$, in 
Fig. 9, the contribution under consideration vanishes. 

This proves that the contribution of Fig. 9 to the production rate 
is nonleading. 

\newpage 
\begin{center} 
{\Large {\bf Figure captions} } \vspace*{1.5em} 
\end{center} 
\begin{description} 
\item[Fig. 1.] Diagram that yields leading contribution to the 
soft-photon production rate in HTL-resummation scheme. \lq\lq $1$'' 
and \lq\lq $2$'' designate the type of photon-quark vertex. $P$, 
$K$, and $K'$ are soft and the blobs on the solid lines indicate the 
effective quark propagators and the blobs on the vertices indicate 
the effective photon-quark vertices. 
\item[Fig. 2.] HTL of the photon-quark vertex. \lq\lq $\ell$'', 
\lq\lq $i$'', and \lq\lq $j$'' are thermal indices. 
\item[Fig. 3.] Thermal self-energy part of a hard quark. 
\item[Fig. 4.] \lq\lq Modified'' HTL of the photon-quark vertex. 
The square blobs on the solid lines indicate self-energy-part 
resummed hard-quark propagators, $\displaystyle{ 
\raisebox{1.1ex}{\scriptsize{$\diamond$}}} \mbox{\hspace{-0.33ex}} 
S$s. 
\item[Fig. 5.] \lq\lq Modified'' HTL of the quark self-energy part. 
The square blob on the curly line indicates $\displaystyle{ 
\raisebox{1.1ex}{\scriptsize{$\diamond$}}} \mbox{\hspace{-0.33ex}} 
\Delta^{\mu \nu} (Q)$. 
\item[Fig. 6.] An $n$-loop ladder diagram for the photon-quark 
vertex function. Solid lines stand for quarks and dashed lines stand 
for gluons. 
\item[Fig. 7.] Nonladder diagram for the photon-quark vertex 
function. 
\item[Fig. 8.] A \lq\lq correction'' to the HTL of the photon-quark 
vertex. The square blob on the vertex indicates $\hat{\Lambda}^\mu 
(Q + K', Q + K)$. 
\item[Fig. 9.] A two-loop contribution to a photon-quark vertex. 
\end{description} 
\end{document}